\documentclass[prd,aps,floats,floatfix,eqsecnum,nofootinbib,12pt]{revtex4-1}
\usepackage{verbatim,graphicx,amssymb,amsbsy,bm,amsmath,rotating,epsfig}
\usepackage{psfrag}
\usepackage{hyperref}
\usepackage{fontenc}
\usepackage[latin1]{inputenc}
\usepackage{pstricks,pst-node,pst-text,pst-3d}
\usepackage{textcomp}
\newcommand{\be}{\begin{equation}}
\newcommand{\ee}{\end{equation}}
\newcommand{\bea}{\begin{eqnarray}}
\newcommand{\eea}{\end{eqnarray}}

\begin{document}

\title{New Quantum Phase of the Universe before Inflation
and its Cosmological and Dark Energy Implications}
\author{Norma G. SANCHEZ\\ 
CNRS LERMA Observatoire de Paris PSL University,\\
Sorbonne Universit\'e, 
61, Avenue de l'Observatoire, 75014 Paris, France}
\date{\today }

\begin{abstract}
{\bf Abstract:} The physical history of the Universe is completed by including the quantum planckian  and super-planckian phase before Inflation in the Standard Model of the Universe in agreement with observations. 
In the absence of a complete quantum 
theory of gravity, we start from quantum physics and its foundational milestone: The {\it universal} classical-quantum (or wave-particle) duality, 
which we extend to gravity and the Planck domain. As a consequence, classical, quantum planckian and superplanckian regimes are covered, and the usual
quantum domaine as well. A new quantum precursor phase of the Universe appears beyond the Planck scale ($t_P$): $10^{-61} t_P \leq t \leq t_P$; the known classical/semiclassical Universe being in the range: $t_P \leq t \leq 10^{+61} t_P$.
 We extend in this way de Sitter universe to the quantum domain: {\it classical-quantum de Sitter duality}. As a  result: (i) The classical and quantum dual de Sitter Temperatures 
and Entropies are naturally included, and the different (classical, semiclassical, 
quantum planckian and super-planckian) de Sitter regimes characterized in a precise and unifying way. 
(ii) We apply it to relevant cosmological examples as the CMB, Inflation 
and Dark Energy. This allows to find in a simple and consistent way: 
(iii) Full quantum Inflationary spectra and their CMB observables, including in particular the 
classical known Inflation spectra and the quantum corrections to them. (iv) A whole unifying picture for the Universe epochs and their quantum precursors 
emerges with the cosmological constant as the vacuum energy, entropy and temperature of the Universe, clarifying the so called cosmological constant problem 
which once more in its rich history needed to be revised.\\
Norma.Sanchez@obspm.fr,  
\end{abstract}


\maketitle
\tableofcontents

\section{Introduction and Results}

The set of robust cosmological data (cosmic microwave background, large scale
structure and deep galaxy surveys, supernovae observations, measurements of the Hubble-Lemaitre constant and other data) support the Standard (concordance) Model of the Universe and place de Sitter (and quasi-de Sitter) stages as a real part of it \cite{WMAP1},\cite{WMAP2},\cite{Riess},\cite{Perlmutter},\cite{Schmidt}\cite{DES},\cite{Planck6}. Moreover, the physical classical, semiclassical and quantum planckian and super-planckian de Sitter regimes are particularly important for several reasons:

\medskip

{\bf (i)} The classical, present time accelerated expansion of the Universe and its associated dark energy or
 cosmological constant in the today era: classical cosmological de Sitter regime.

\medskip

{\bf (ii)} The semiclassical early accelerated expansion of the Universe and its associated Inflation era: semiclassical cosmological de Sitter (or quasi de Sitter) regime (classical general relativity plus
quantum field fluctuations.)
\medskip

{\bf (iii)} The quantum, very early stage preceeding the Inflation era: Planckian and super-Planckian 
quantum era. Besides its high conceptual and fundamental physics interest, this era could be of realistic 
cosmological interest for the test of quantum theory itself at such extreme scales, as well as for the 
search of gravitational wave signals from quantum gravity for e-LISA \cite{LISA} for instance, after the 
success of LIGO \cite{LIGO},\cite{DESLIGO}. In addition, this quantum stage should be relevant in 
providing quantum precursors and consistent initial states for the semiclassical (fast-roll and slow 
roll) inflation, and their imprint on the observable primordial fluctuation spectra for instance. 
Moreover, and as a novel result of this paper, this quantum era allows to clarify the issue of dark 
energy as the vacuum energy or cosmological constant of the Universe.

\medskip

{\bf (iv)} de Sitter is a simple and smooth constant curvature vacuum background without any physical 
singularity, it is  maximally symmetric and can be described as a hyperboloid embedded in Minkowski space-
time with one more spatial dimension. Its radius, curvature and equivalent density are described in terms of only 
one physical parameter: the cosmological constant.

\medskip

In despite of the simplicity of de Sitter background, the generic nature of inflation, and the relevance 
of dark energy, there is no satisfactory description of de Sitter 
background, nor of inflation in string theory, -  it is fair to recall here that this was pointed out  
25 years ago \cite{deVega1994}. [Contrary to Anti-de Sitter, de Sitter background does not appear as a solution of 
the effective string equations. The lack of a full conformal invariant string de Sitter description is 
not an handicap for de Sitter background, but to the current formulation or understanding of string 
theory, \cite{deVega1993} (by satisfactory description we mean in particular, one not based in tailored 
constructions, nor in conjectures)]. 

\medskip

The lack of a complete theory of quantum gravity (in field theory and in strings) 
does not preclude to explore and describe planckian and superplanckian 
gravity regimes.
Instead of going from classical gravity to quantum gravity by quantizing general relativity (as it was 
tried with its well known developpements and shortcomings, (is not our aim here to review it), we start 
from Quantum physics and its foundational milestone:
the classical-quantum (wave-particle) duality, and extend it to include gravity and the Planck scale 
domain, namely,  wave-particle-gravity duality,  
(or classical-quantum gravity duality), \cite{Sanchez2019}, \cite{Sanchez2003-1}. As a consequence, the different gravity regimes are covered: 
classical, semiclassical and quantum, together with the Planck domain and the elementary particle  
domain as well. This duality is {\it universal}, it includes the known classical-quantum duality as a 
special case and allows a general clarification from which physical understanding and cosmological 
results can be extracted as shown in this paper. This is not an assumed or conjectured duality. As the 
wave-particle duality, this does not rely 
on the number of space-time dimensions (compactified or not), nor on any symmetry or isometry nor on any 
other {\it at priori} condition.
 
\medskip

{\bf In this paper}, we link the Standard Model of the Universe to the classical-quantum duality. We complete in this way the history of the Universe beyond the Inflation era and the current picture by including the quantum precursor phase within the Standard Model of the Universe in agreement with observations. Quantum physics is more complete than classical physics 
and contains it as a particular case: It adds a new quantum planckian and 
superplanckian phase of the Universe from the Planck time $t_P$ untill the extreme past $10^{-61} t_P$ which is an upper bound for the origin of the Universe, 
with energy  $H = 10^{61} h_P$. Besides the arguments given above, the reasons supporting such a phase are many: (i) The generic and physical existence of classical-quantum duality in Nature.
(ii) The universality of it. 
(iii) The consistent and concordant physical results and coherent whole picture obtained from it supported by observations.
We provide an unifying description of the classical, semiclassical, quantum, 
planckian and superplanckian stages of the Universe, their relevant physical 
magnitudes: size, mass, vacuum energy density, cosmological constant, gravitational entropy  and temperature  and the relations between them. 

{\bf The main results of this paper:}
 {\bf 1.} The classical dilute Universe today and the highly dense very early quantum superplanckian Universe are classical-quantum duals of each other in the precise meaning of the classical-quantum duality. This means the following:
The classical Universe today $U_{\Lambda}$ is clearly characterized by the set of physical gravitational 
magnitudes or observables (age or size,  mass, density, temperature, entropy) $\equiv (L_\Lambda, M_\Lambda, \rho_\Lambda, T_\Lambda, S_\Lambda)$
\begin{equation}\label{ULambda1}
U_{\Lambda} = (L_\Lambda, M_\Lambda, \rho_\Lambda, T_\Lambda, S_\Lambda)
\end{equation}
The highly dense very early quantum Universe $U_Q$ is characterized by the corresponding set of quantum dual physical quantities $(L_Q, M_Q, \rho_Q, T_Q, S_Q)$ in the precise meaning of the classical-quantum duality:
\begin{equation} \label{UQ1}
U_Q = (L_Q, M_Q, \rho_Q, T_Q, S_Q)
\end{equation}
\begin{equation} \label{Udual1}
U_Q =  \frac{u_P^2}{U_\Lambda}, \qquad u_P = (l_P, m_P, \rho_P, t_P, s_P)
\end{equation} 

$u_P$ standing for the corresponding quantities at the fundamental constant Planck scale,
the {\it crossing scale} between the two main (classical and quantum) gravity domains.
The classical $U_{\Lambda}$ and quantum $U_Q$ Universe eras or regimes (classical/semiclassical eras of the known Universe and its quantum planckian and superplanckian very early phases), satisfy Eqs.(\ref{ULambda1})-(\ref{Udual1}). The {\it total} Universe $U_{Q\Lambda}$ is composed by their classical/semiclassical and quantum phases:
\begin{equation} \label{Utotal1}
U_{Q\Lambda} = \left(\; U_Q  +  U_{\Lambda} + u_P \;\right)
\end{equation}
$Q$ stands for Quantum, $\Lambda$ for classical, and $P$ for the fundamental Planck scale 
constant values.

\medskip

In particular, the quantum dual de Sitter universe $U_Q$ is generated  
from the classical de Sitter universe  
$U_{\Lambda}$ through Eqs.(\ref{ULambda1})-(\ref{Utotal1}): {\it classical-quantum de Sitter 
duality}, as we done in Section III. The {\it total} (classical plus quantum dual) de Sitter 
universe $U_{Q\Lambda}$ is obtained in Section IV: {\it classical-quantum de Sitter symmetry}. 
This includes in particular the classical, quantum and total de Sitter Temperatures and Entropies 
(Sections V and VI). This allows to characterize in a complete and  precise way the different 
classical, semiclassical, quantum planckian and superplanckian de Sitter regimes (Section VII). 
$H$ stands for the classical Hubble-Lemaitre constant, or its equivalent $\Lambda = 3 \; \left
({H}/{c}\right)^2$. $Q$ stands for quantum dual, and $QH$ (or $Q\Lambda$) for the total or complete quantities.

\medskip

For instance, the size of the Universe is the gravitational length $L_\Lambda = \sqrt{ 3/\Lambda}$
in the classical regime, it is the quantum Compton length $L_Q$ in the quantum dual regime (which is the full quantum planckian and superplanckian regime), and it is the Planck length $l_P$ at the fundamental Planck scale: the {\it crossing scale}. The {\it total}  (or complete) size $L_{Q \Lambda}$ is the sum of the two components. Similarly, the horizon acceleration (surface gravity) $K_{\Lambda}$ of the Universe in its classical gravity regime becomes the quantum acceleration $K_Q$ in the quantum dual gravity regime. The temperature $T_{\Lambda}$, measure of the classical gravitational length or mass becomes the quantum Temperature $T_Q$ (measure of the quantum size or Compton length) in the quantum  regime. Consistently, the  Gibbons-Hawking temperature  
is {\it precisely} the quantum temperature $T_Q$. Similarly,  the classical/semiclassical gravitational area or entropy $S_\Lambda$ (Gibbons-Hawking entropy) has its quantum dual $S_Q$ in the quantum gravity (Planckian and super-Planckian) regime. The concept of gravitational entropy is {\it the same} for any of the gravity regimes: $ Area / 4 l_P^2$ in units of $k_B$.
(For a classical object of size $L_\Lambda$, this is the classical area $A_\Lambda$, for a quantum object, of size $L_Q$, this is the area $A_Q$.)

\medskip

{\bf 2. Results for Inflation.} We apply these results to Inflation, Dark Energy and the cosmological constant in the framework of the Standard Model of the Universe, (Sections VIII-XI). The precursor quantum phase of the known classical/semiclassical Inflation does appear, as well as the precursors for the classical standard eras and today Dark Energy era. H-inflation means the classical known Inflation (classical H) era, Q-inflation is its quantum dual precursor, QH stands for the total Inflation era including the known classical/semiclassical Inflation and its precursor: the quantum Inflation era (in  the planckian and superplanckian) phase. The {\it total} or complete QH
inflationary spectra turn out expressed as
\begin{equation}
[\;\Delta^S_{k, \;QH}\;] = [\;\Delta^S_{k, \;H}\;]\; \left(\frac{1}{[ \;1 + (H/h_P)^2 \;] }\right)\; 
\frac{1}{( 1 - \delta \epsilon_{QH})^{1/2}}
\end{equation}
\begin{equation}
[\;\Delta^T_{k, \;QH}\;] = [\;\Delta^T_{k, \;H}\;] \; \left(\frac{1}{[\; 1 + (H/h_P)^2 \;] }\right)
\end{equation}

\medskip

where $[\Delta^S_{k,\;H}]$ and  $[\Delta^T_{k, \;H}]$ are the known standard spectra of scalar 
curvature and tensor perturbations in classical H Inflation Eqs.(\ref{DeltaSO}). Here $\delta\epsilon_{QH}$ is the first order QH slow-roll parameter (computed in Section IX) which contain in particular the classical known slow-roll $\epsilon$ parameter, and $h_P$ is the value of the Hubble constant at the Planck scale (or mass Planck value $m_P$). The total QH spectra contain both: the standard known spectra of the classical/
semiclassical Inflation including its quantum corrections of order $(H/h_P)^2 = 10^{-12}$ in the classical/semiclassical gravity phase $H = 10^{-6} h_P$, at $t = 10^{6} t_P$, (or $10^{-5} M_P$ for the reduced Planck mass $M_P = m_P/\sqrt{8\pi}$), 
and their quantum dual spectra in the quantum precursor Inflation era $H_Q = 10^6 h_P$, at $t = 10^{-6} t_P$. 

The CMB observables: scalar spectral index $n_S$, ratio $r$ and departure from scalar 
invariance $\Delta$ are computed (in the two Inflation phases, classical H, quantum Q, and the total QH): In 
the classical H known phase, it yields in a simple and direct way the same quantum corrections to the spectra, sign and magnitude, as the quantum inflaton corrections \cite {Boya2005},
\cite{Boya2006} in the Ginsburg-Landau 
effective approach to Inflation \cite{CiridVS},\cite{BDdVS}.
The {\it departure from scale invariance} $\Delta_{QH} = (n_{s\;QH} - 1)/2 + r_{QH}/8$ ,
gets corrected as [Eq.(\ref{DeltaQH2})]: $$ \Delta _{QH} = \Delta\;[\; 1 - 2 \; (H/h_P)^2 \;] + \sqrt{\epsilon} \;(H/h_P)^2
\; [\;2 \sqrt{\epsilon} - \frac{m_P}{\sqrt{\pi}}\;] + O\;(H/h_P)^4,$$
where $\epsilon$ and $\Delta = (\;n_s - 1\;)/2 + r/8 $ are the  slow roll parameter and 
the {\it departure from scale invariance} of the known classical H Inflation respectively.  
The QH corrections to the known scalar index $n_s$ and ratio $r$ of classical/semiclassical Inflation $ (H/h_P) = 10^{-6}$ are: $$
\frac{r_{QH}}{r} - 1 = - 2 \;10^{-12}, \; \qquad \frac{n_{s\;QH}}{n_s} - 1 =
2 \;10^{-12}\;[\;1 - \frac{1}{n_s} (\;1- \frac{m_P}{2}\sqrt{\frac{\epsilon}{\pi}}\;)\;].$$
The QH factor modifying the Hubble constant and the complete QH inflation spectra:
\begin{equation}\label{sum1}
QH \equiv \frac{H}{[\;1 + (H / h_P)^2\;]} = \frac{H_Q}{[\;1 + (H_Q / h_P)^2\;]} =  
H \; \sum _{n=0}^{\infty} (-1)^n \left(\frac{H}{h_P}\right)^2,
\end{equation}
covers the {\it full classical and quantum range} 
$H \leq h_P$ and $H \geq h_P$. If $H< h_P$, it covers the known classical/semiclassical  range. If $H > h_P$, it changes consistently to the quantum Hubble rate $H_Q = h_P^2/H$ in the quantum domain.

\medskip

{\bf 3. Results for Dark Energy.} This framework reveals enlighting for the issue of {\it Dark Energy} as  discussed in Section X, and allows clarification into the cosmological constant problem as discussed in Section XI. The classical Universe today $U_\Lambda$ is precisely  a {\it classical dilute gravity vacuum dominated by voids and supervoids} as shown by observations \cite{VoidsHistory},
\cite{Voids1}, \cite{VoidsPRL} whose observed $\rho_\Lambda$ or $\Lambda$ value today \cite{Riess},\cite{Perlmutter},\cite{Schmidt},\cite{DES},\cite{Planck6}
is {\it precisely} 
the classical dual of its quantum precursor values $\rho_Q,\Lambda_Q$ in the quantum very early precursor vacuum $U_Q$ as determined by Eqs.(\ref{ULambda1})-(\ref{UQ1}). The high density $\rho_Q$ and cosmological constant $\Lambda_Q$ are 
precisely the quantum particle physics superplanckian value $10^{122}$. This is precisely expressed by Eqs.(\ref{ULambda1})-(\ref{UQ1}) applied to this case, Section X [Eqs.(\ref{LambdaHvalue})-(\ref{LambdaQLambda})]: 
\begin{equation} \label{LambdaHvalue1} 
\Lambda = 3 H^2 = \lambda_P  \left (\frac{H}{h_P}\right)^2 = \lambda_P \left (\frac{l_P}{L_H}\right)^2
= (2.846 \pm 0.076) \; 10^{-122}\; m_P^2
\end{equation}
\begin{equation} \label{LambdaQvalue1} 
\Lambda_Q = 3 H_Q^2 = \lambda_P \left (\frac{h_P}{H}\right)^2 = \lambda_P \left (\frac{L_H}{l_P}\right)^2
= (0.3516 \pm 0.094) \; 10^{122}\;h_P^2 
\end{equation}
\begin{equation} \label{LambdaQLambda1} 
\Lambda_Q = \frac{\lambda_P^2}{\Lambda}, \qquad \lambda_P = 3 h_P^2
\end{equation}
{\it The quantum dual value $\Lambda_Q$ is {\it precisely} the quantum vacuum value $\rho_Q = 10^{122}\; \rho_P$ obtained from particle physics:}
\begin{equation} \label{rhoQii1}
\rho_Q = \rho_P \left(\frac{\Lambda_Q}{\lambda_P}\right) 
= \frac{\rho_P^2}{\rho_\Lambda} = 10^{122}\; \rho_P
\end{equation}

Eqs.(\ref{LambdaHvalue1})-(\ref{rhoQii1}) are consistently supported by the data \cite{Riess},\cite{Perlmutter},\cite{Schmidt},\cite{DES},\cite{Planck6} which we also {\it link to the gravitational entropy and temperature of the Universe}, as we done in Section XI and summarized by Eqs.(\ref{Lambda1}) - (\ref{LambdaQLambdavalue}). 
The  {\it complete} cosmological constant $\Lambda_{Q\Lambda}$ or total vacuum energy density 
$\rho_{Q\Lambda}$ is the sum of its classical and quantum components (corresponding to the classical today era and its quantum  planckian and super-planckian precursor):
\begin{equation} \label{LambdaQLambda3}
\Lambda_{Q \Lambda} = \lambda_P \left (\;    
  \frac{\Lambda}{\lambda_P} + \frac{\lambda_P}{\Lambda} + 1\;\right) = 
\lambda_P \;(\; 10^{-122} + 10^{+122} + 1\;)
\end{equation}

The observed $\Lambda$ or $\rho_\Lambda$ today is the {\it classical gravity vacuum} value 
of the classical Universe $ U_\Lambda$ {\it today}. Such observed value must be 
consistently in such way because of the {\it large classical} size of the Universe today  $L_
\Lambda = \sqrt{3/\Lambda}$, and of the empty or vacuum  dilute state today dominated by {\it 
voids and supervoids} as shown by the set of large structure observations \cite{VoidsHistory},
\cite{Voids1}, \cite{VoidsPRL}. This is one main physical reason for such a {\it low} $\Lambda$ 
value at the present age today  $10^{61} t_P$. Its precursor value and 
density $\Lambda_Q , \rho_Q$ is a high superplanckian value precisely because this is a high density 
very early {\it quantum cosmological vacuum} in the extreme past $10^{-61} t_P$ of the quantum superplanckian precursor phase $U_Q$.

\medskip

The quantum cosmological  constant and associated density  
$\Lambda_Q = \rho_Q = 10^{122}$ (in Planck units) in the quantum precursor superplanckian phase $U_Q$ at $10^{-61} t_P$, (the extreme past), became  
the classical cosmological constant and density $\Lambda = \rho_\Lambda = 10^{-122}$ 
in the classical Universe $U_\Lambda$ today at $10^{61} t_P$. 
The superplanckian value is consistently in such way because is a extreme quantum gravity vacuum 
in the extreme quantum past $10^{-61} t_P$ with minimal entropy $S_Q = 10^{-122} = \Lambda = \rho_\Lambda$.
 All physical quantities: the vacuum energy density, the 
cosmological constant, the gravitational entropy and gravitational temperature, both classical and 
quantum are consistently linked by the {\it classical-quantum (or wave-particle) duality through the Planck scale} in agreement with observations Eqs.(\ref{ULambda1})-(\ref{Udual1}), (Sections X-XI).

\medskip

Eqs.(\ref{LambdaHvalue1}) to (\ref{rhoQii1}),(\ref{LambdaQLambda3}), [Eqs.(\ref{Lambda1})-(\ref{LambdaQLambdavalue})]  
concisely and synthetically express such complete set of classical-quantum dual relations and {\it explain  why} 
the classical gravitational vacuum: cosmological constant $\Lambda$ or density $\rho_\Lambda$ {\it coincides} with such observed {\it low value} $10^{-122}$ in Planck units, and {\it why} their corresponding quantum gravity precursor vacuum has such extremely {\it high} superplanckian density {\it value} $10^{122}$ in Planck units. This is 
{\it not} trivial, this is simple, deep and robust, (Section X). This is {\it not} a tailored argument or construction to the dark energy /cosmological constant problem. This is a consequence of a whole general picture, (Section XI), within the Standard Model of the Universe.

\medskip

{\bf 4. A Whole picture.} Overall, a consistent unifying clear picture of the history of the Universe does emerge in terms of the gravitational classical, semiclassical and quantum phases and their relevant characterizing physical magnitudes as the size, age, vacuum density,  gravitational entropy and temperature, all in terms of the cosmological constant. This sheds light in the Inflation and Dark energy eras and in the cosmological constant problem. This is summarized in the end of Section XI (the whole history), and depicted in Fig.(1).

\medskip

The evolution of the Universe can be described by two big phases: Classical and Quantum, that is to say,  after 
and before the Planck time $t_P = 10^{-44}\; sec$ respectively. Each cosmological stage in the classical known Universe $t_P \leq t \leq 10^{61}\;t_P$ has a dual quantum stage in the preceding quantum phase 
before the Planck time: $10^{-61}\; t_P \leq t \leq t_P$. 

\medskip

The whole duration (of the classical plus quantum phases) is precisely
$10^{-61}\; t_P \leq t \leq 10^{+61}\; t_P$. That is to say, 
{\it each} component {\it naturally} dominates in each phase: classical time 
component $ 10^{+61}\; t_P $ in the classical era, quantum Planck value $t_P$ 
in the quantum preceding era.  

\medskip

The present time of the Universe at $10^{+61}\; t_P$, which is {\it a lower bound} for the future (if any) age of the Universe, has a remote past quantum precursor equal to $10^{-61}\; t_P$, which is an {\it upper bound} for the origin of the Universe. The classical/semi-classical known inflation era which occurred at about $10^{+6}\; t_P, H = 10^{-6}\; h_P$ has a preceding quantum dual era at $10^{-6}\; t_P,  H = 10^{6}\; h_P$ which is a semi-quantum era ('low $H$' with respect to the extreme past quantum state $H = 10^{61}\; h_P$), and similarly, for any of the other known eras in the Classical post-planckian Universe. This appears to be the way in which the Universe has evolved. A complete picture is discussed in Section XI including the gravitational entropy and temperatures, and summarized in Fig.(1).

\medskip

In Planck units, is the same to express the age of the Universe $10^{61}$  in terms of time, length, mass, temperature  or square root of entropy (arrow of time) to describe the complete Universe. Similarly, the vacuum energy density is the dual to the gravitational entropy. The complete quantum theory is a theory of {\it pure numbers}.

\medskip

The {\it total or complete} physical quantities are invariant under the classical-quantum duality: 
$H \leftrightarrow Q$, as it must be: This means physically that: (i) what occurred in the quantum phase 
before $t_P$ {\it determines} through quantum duality Eqs.(\ref{ULambda1})-(\ref{Utotal1}) what occurred in the classical phase after $t_P$. And: (ii) what occurred in the quantum phase before $t_P$ is the {\it same physical observable}, or event which occurred after $t_P$ in the precise meaning of the classical-quantum dual relations 
Eqs.(\ref{ULambda1})-(\ref{Udual1}). That is to say: The quantum dual quantities in the quantum phase before 
$t_P$, are the {\it quantum precursors} of the classical/semiclassical quantities after $t_P$. 
As the wave-particle duality at the basis of quantum mechanics, the wave-particle-gravity 
duality, 
is reflected in all cosmological eras and its associated quantites, temperatures and 
entropies as well.

\medskip

Cosmological evolution goes from a quantum planckian and superplanckian phase to a 
semiclassical accelerated era (de Sitter inflation), then to the classical 
 eras untill the present classical de Sitter phase. 
The classical-quantum or wave-particle-gravity duality specifically manifests
in this evolution, between the different gravity regimes, and could be view 
as a mapping between asymptotic (in and out) states characterized by sets 
$U_Q$ and $U_\Lambda$ and thus as a Scattering-matrix description.

\medskip

{\bf This paper is organized as follows:} In
Section II we describe the classical - quantum duality including gravity 
together with its properties covering the different gravity regimes:
(classical, semiclassical and quantum gravity domains) passing through the Planck scale 
and the elementary particle domain as well. 
In Sections III and IV we describe the classical, quantum dual and  
complete de Sitter universe, its physical duality symmetry 
and its properties.
Sections V and VI deal with the classical, quantum dual and complete de Sitter Temperature and 
Entropy. In Section VII we characterize in a precise and unifying way the different (classical, 
semiclassical, Planckian and super-Planckian) de Sitter regimes. Sections VIII-X illustrate 
the results with relevant cosmological examples and their implications for 
the CMB fluctuations, Inflation and Dark Energy. In Section XI we provide a clarifying 
unifying picture with the cosmological constant as the vacuum energy, entropy and temperature 
of the Universe. Section XII summarizes outlook and conclusions.

\section{Classical - Quantum Duality through the Planck scale}

Let us stand by $O_{G}$ the set of relevant physical variables or observables characteristic of the 
classical gravity regime, (as size, mass, surface gravity (or gravity acceleration), and  
usual temperature for instance), and by $O_Q$ the corresponding set 
of quantities in the \textbf {quantum dual}
regime in the precise sense of the wave-particle or classical-quantum duality: 
The magnitudes $O_{G}$ and $O_Q$ are classical-quantum gravity duals of each other, 
(in the precise meaning  of the classical-quantum duality) here through the Planck scale, \cite{Sanchez2019}:
\begin{equation} \label{O}
O_{G}=o_{P}^2\, O_Q^{-1}
\end{equation}

This relation holds in general for any quantity in the set. $O_{G}$ and 
$O_Q$ are the same conceptual physical quantities in the different 
(classical/semiclassical and quantum) gravity regimes respectively. 
The constant $o_{P}$ stands for the corresponding quantity at the Planck scale, ie purely 
depending of the fundamental constants ($\hbar ,\, c,\, G)$. 
$O_Q$ stands for relevant quantum concepts 
as quantum size $ L_Q$, quantum mass $ M_Q$, quantum acceleration $ K_Q = c^2/L_Q$, 
 quantum temperature $T_Q$ and other physical magnitudes associated to them .

\medskip

This is not an assumed or conjectured duality. This duality is {\it universal}.
As the wave-particle duality, 
this classical/semiclassical-quantum gravity duality does not relate to the number of 
dimensions, nor to any particular or imposed symmetry of the background manifold or space- 
time, nor to any other condition.

\medskip

Each of the sides of Eq.(\ref{O}) accounts for each domain separately: 
classical {\it or} quantum, ie $ O_G$ {\it or} $O_Q$, and their respective associated 
set of magnitudes. The 
{\it total or complete} or QG magnitudes take into 
account the different gravity domains: classical and quantum, and their duality properties, 
passing through the Planck scale and including the elementary particle domain as well \cite{Sanchez2019}:
\begin{equation} \label{OQG}
O_{QG} = (O_Q + O_G)
\end{equation}
In Planck units, the complete QG magnitudes simply read
\begin{equation} \label{Oop}
O_{QG} = o_P \; ( o + \frac{1}{o} ), \qquad o \equiv \frac{O_G}{o_P} = \frac{o_P}{O_Q}
\end{equation}
The two domains $(O\geq o_P)$ and $(O \leq o_P)$ being the classical and quantum domains 
respectively, with the two ways of reaching the Planck scale.

\medskip

The QG magnitudes cover all the classical and quantum domains, with and without gravity. The two  domains precisely account for the elementary particle domain: $0 \leq O \leq o_P $,
and for the macroscopic gravity domain:  $o_P \leq  O\leq \infty$. 
These two domains are duals of each other in the precise sense of 
the classical-quantum duality through the  Planck scale: we call it "Planck scale duality". 
For instance:
Quantum particle theory has $L_Q >> l_P$ \quad and \quad $L_G << l_P$.    
Classical gravity has $L_Q << l_P$ \quad and \quad $L_G >> l_P$.
Quantum Gravity has $L_{QG}$  and any value of $ L_G$ and $L_Q$, 
and includes the Planck domain as well. We implement the classical-quantum duality in de Sitter 
universe in the next section.

\section{Classical and Quantum Dual de Sitter Universes}

de Sitter space-time in $D$ space-time dimensions is the hyperboloid
embedded in Minkowski space-time of $(D+1)$ dimensions:
\begin{equation} \label{dS}
X^2 - T^2 + X_j X^j + Z^2 = L_H^2, \qquad j= 2, 3, ...(D-2)
\end{equation}
$L_H$ is the radius or characteristic length of the de Sitter universe. 
The scalar curvature $ R$ is constant. Classically: 
$$
L_H = {c}/{H}, \qquad R =  H^2 D(D-1) = \frac{2D}{(D-2]}\;\Lambda, 
\qquad \Lambda = \frac{H^2}{2} (D-1) (D-2)
$$

\bigskip

Moreover, a mass $M_H$ can be associated to $L_H$ or $H$, such that (we take D = 4 here for simplicity):
\begin{equation} \label{LH}
L_H = \frac{G M_H}{c^2} \equiv L_G, \;\;\qquad M_H = \frac{c^3}{GH} 
\end{equation}
The corresponding quantum magnitudes $L_Q$, $M_Q$ are the quantum  duals of 
$L_H$, $M_H$ respectively
in the precise meaning of the classical-quantum (de Broglie or Compton) duality:
\begin{equation} \label{MH}
L_Q = \frac{\hbar}{M_H c} = \frac{\hbar\; G H}{c^3}\;, \;\;\qquad M_Q = \frac{\hbar H}{c^2}
\end{equation}
\begin{equation} \label{MQ}
{\text ie,} \quad  L_Q = \frac{l_P^2}{L_H}\;, \;\;  \qquad M_Q = \frac{m_P^2}{M_H}
\end{equation}
where $l_P$ and $m_P$ are the Planck length and Planck mass respectively:
\begin{equation}\label{lp}
l_P = \sqrt{\frac{\hbar\;G}{c^3}}\;, \;\; \qquad m_P = \sqrt{\frac{c\;\hbar}{G}}
\end{equation}

Similarly, for the quantum dual Hubble constant  $H_Q$ and the quantum curvature $R_Q$:
\begin{equation}\label{dualQH}
H_Q = \frac{h_P^2}{H} , \qquad R_Q = \frac{r_P^2}{R}, 
\qquad \Lambda_Q = \frac{\lambda_P^2}{\Lambda}
\end{equation}  

where $ h_P, r_P, \lambda_P $ are the Planck scale values of the Hubble constant, 
scalar curvature and cosmological constant respectively:
\begin{equation}\label{P}
h_P = \frac{c}{l_P} ,\qquad r_P = h_P^2 \;D (D-1) 
, \qquad \lambda_P = \frac{h_P^2}{2} \;(D-1) (D-2) 
\end{equation}  
\begin{equation}\label{PD4}
h_P = c^2 \sqrt{\frac{c}{\hbar G}},\qquad r_P =  12 \; h_P^ 2 = 4 \;\lambda_P ,\qquad 
\lambda_P = 3 \; \left(\frac{c^5}{\hbar G}\right),  \qquad (D=4)
\end{equation}

\section {Total de Sitter Universe and its Duality Symmetry}

The classical $L_H \equiv L_G$ length and the quantum length $ L_Q$ can be extended 
to a more complete length $L_{QH}$ which includes both: 
(we call it {\it complete} or Quantum Gravity (here QH) length since it contains both: Q and H lengths):
\begin{equation}\label{LQH}
L_{QH} =  (L_H + L_Q) = l_P \; (\frac{L_H}{l_P} + \frac{l_P}{L_H}). 
\end{equation}
and we have then : 
\begin{equation}\label{ZQH}
X^2 - T^2 + X_j X^j + Z^2 = L_{QH}^2 = 
 2\;l_P^2 \left[1 + \frac{1}{2}\; [\;(\frac{L_H}{l_P})^2 + (\frac{l_P}{L_H})^2\;]\right]
\end{equation}
with   $j = 2,3, ...(D-3)$. 

\medskip

Eq.(\ref{ZQH}) quantum generalize de Sitter space-time including the classical, 
semiclassical and quantum de Sitter regimes and the Planck scale de Sitter regimes as well. 
It contains two non-zero lengths $(L_H, L_Q)$ or two relevant scales ($H$,  $l_P$) 
enlarging the possibilities for the space-time regimes or phases: 
Quantum, semiclassical and classical de Sitter regimes. Thus,

\medskip

\begin{itemize}
\item{For $ L_H >> l_P$, ie $L_Q << L_H$, Eq.(\ref {ZQH}) yields the classical de Sitter space-time. 
For intermediate $L_H$ values between $l_P$ and $L_Q $ it yields
the semiclassical de Sitter space-time. }

\item{For $ L_H = l_P $ ie $L_Q =  l_P = L_{QH}$, Eq.(\ref{ZQH}) yields
 the Planck scale de Sitter hyperboloid.}
\item{For $ L_H << l_P $, ie $L_Q >> L_H $ it yields the highly quantum de Sitter regime, deep 
inside the Planck domain.}
\end{itemize}

\medskip

$H = c / L_H $ is ($ c ^{-1} )$ times the surface gravity (or gravity acceleration) 
of the classical de Sitter space-time. Similarly, $ H_Q = c / L_Q $ 
and $H_{QH} = c / L_{QH}$ are the surface gravity in the quantum and whole  
QH de Sitter phases respectively.

\medskip

Similarly, from Eq. (\ref{LQH}) and Eqs (\ref{LH})-(\ref{MQ}), we have for the mass:
\begin{equation}\label{MQH}
M_{QH} =  (\;M_H + M_Q\;) = m_P \;(\;\frac{M_H}{m_P} + \frac{m_P}{M_H}\;)
\end{equation}
\begin{equation}\label{MLQH}
\frac{M_{QH}} {m_P} = m_P \;(\;\frac{L_H}{l_P} + \frac{l_P}{L_H}\;) = \frac{L_{QH}}{l_P}
\end{equation}

$M_{QH}/m_P$ and  $L_{QH}/l_P$ both have the same expression with respect to their
respective Planck values.

\bigskip

{\bf The complete QH Hubble constant $H_{QH}$, curvature $R_{QH}$ and $\Lambda_{QH}$.} 
The fully quantum  QH Hubble $H_{QH}$ constant, curvature $R_{QH}$ and $\Lambda_{QH}$ constant
follow from the QH de Sitter length $L_{QH}$ Eq.(\ref{LQH}):
\begin{equation}       \label{QH1}
H_{QH} = \frac{c}{L_{QH}}, \qquad  R_{QH} = H_{QH}^2\; D\;(D-1), 
\qquad  \Lambda_{QH} =  \frac{H_{QH}^2}{2} \;(D-1) (D-2)
\end{equation}
where from Eqs.(\ref{LQH}) and (\ref{dualQH}):
\begin{equation}    \label{QH2}
H_{QH} = \frac{H}{ [\;1 + (l_P H / c)^2\;]}, \qquad H_{QH}/ h_P = \frac{(H/h_P)}{ [\;1 + (H / h_P)^2\;]},
\qquad h_P = c / l_P
\end{equation}

\medskip

We see the {\it symmetry} of $H_{QH}$ under $(H/h_P) \rightarrow (h_P/H)$, 
ie under $ H \rightarrow H_Q = (h_P^2/H)$ :
\begin{equation}    \label{symQH}
H_{QH} (H / h_P) = H_{QH} (h_P / H)
\end{equation}

That is, the classical $H$ and quantum $H_Q$ are classical-quantum duals of each other through the Planck 
scale $h_P$, but the complete or total $H_{QH}$ which contain both of them 
is {\it invariant}.
And similarly, for the  quantum curvature  $R_{QH}$ and cosmological constant $\Lambda_{QH}$ Eq.(\ref{QH1})
derived from them :
\begin{equation}\label{symRQH}
R_{QH}(H/h_P) = R_{QH}(h_P/H),  \qquad  \Lambda_{QH}(H/h_P) = \Lambda_{QH}(h_P/H)
\end{equation}
where:
\begin{equation}\label{R1}
R_{QH} = \frac{R_H}{ [\;1 + R_H /r_P \;]^2}  =  \frac{R_Q}{[\;1 + R_Q /r_P \;]^2},\qquad r_P = 12\;h_P^2
\end{equation}

\begin{equation}\label{L1}
\Lambda_{QH} = \frac{\Lambda_H}{[\;1 + \Lambda_H /\lambda_P \;]^2}  =  
 \frac{\Lambda_Q}{[\;1 + \Lambda_Q /\lambda_P \;]^2},\qquad \lambda_P = 3\;h_P^2
\end{equation}

\bigskip

The classical $ H/h_P << 1$, quantum $ H/ h_P >> 1$ and Planck $H/h_P = 1$ 
regimes are clearly exhibited in the  QH expressions Eqs (\ref{QH1}), Eq.(\ref{QH2}):
\begin{equation}\label{sHP}
H_{QH \;(H << h_P)} =   H \;[\; 1 - (H / h_P)^2 \;] + O\;(H / h_P)^4  
= \frac{c}{L_H} \;[ 1 - (\frac{l_P}{L_H})^2 ] + O\;(\frac{l_P}{L_H})^4
\end{equation}
\begin{equation}\label{HP}
H_{QH} \; (H = h_P)  = \frac{h_P}{2}, \; \; h_P= c/l_P
\end{equation}
\begin{equation}\label{bHP}
H_{QH \; (H >> h_P)} =  (h_P^2/H)\; [ 1 - (h_P/H)^2 ] \;+\; O (h_P/H)^4
 = \frac{ c\;L_H}{l_P^2}\; [\; 1 - \;(\frac{L_H}{l_P})^2 \;] \;+ \;O\;(\frac{L_H}{l_P})^4  
\end{equation}

\medskip

The three above equations show respectively the three different de Sitter phases:
\begin {itemize}
\item {The classical gravity de Sitter universe (with lower curvature  than the Planck scale $r_P
$) {\it outside}
the Planck domain $(l_P < L_H < \infty)$.} 
\item {The Planck curvature de Sitter state  $(R_H = r_P, \;\; L_ H = l_P)$}
\item {The highly quantum or high curvature
($R_H >> r_P$) de Sitter phase {\it inside} the quantum gravity Planck domain ($0 < L_H \leq  
l_P$).}
\end {itemize}

\medskip

Eqs (\ref{sHP})-(\ref{bHP}) show the classical-quantum duality through 
the Planck scale: 
The highly quantum gravity regime 
$H_{QH}\;(H >> h_P)$ entirely expresses in terms of the quantum Hubble constant  $H_Q$,
dual through the Planck scale value $h_P$ to the classical/semiclassical Hubble constant $H$, 
Eq.(\ref{dualQH}). 
Is natural to define here the dimensionless magnitudes: 
\begin{equation}\label{dimless}
{\cal L} \equiv \frac{L_{QH}}{l_P}, \qquad {\cal M}\equiv \frac{M_{QH}}{m_P},
 \qquad  {\cal H} \equiv  \frac{H_{QG}}{h_P}, 
\qquad l\equiv \frac{L_H}{l_P}, \qquad  {\quad h} \equiv \frac{H}{h_P} = l^{-1}
\end{equation} 
Then,  Eqs (\ref{LQH}),(\ref{MQH}) and (\ref{QH2}) simply reads:
\begin{equation}\label{dimless2}
{\cal L} =  (l + \frac{1}{l}) = {\cal M},\qquad 
{\cal H} = \frac{1}{(l + \frac{1}{l})} = {\cal L}^{-1}
\end{equation} 
Similarly, for $R_{QH}/r_P$ and $\Lambda_{QH}/\lambda_P$:
\begin{equation}\label{dimlessR}
\frac{R_{QH}}{R_P} = \frac{\Lambda_{QH}}{\Lambda_P} = \left(\frac{H_{QH}}{h_P}\right)^2
\equiv {\cal H}^2
= \frac{1}{(h + h^{-1})^2}
\end{equation}  
In dimensionless variables, the duality symmetry endowed by $L_{QH},M_{QH}$ and Eqs (\ref{symQH}), 
(\ref{symRQH}), simply reads: 
\begin{equation}\label{sym}
{\cal L} (l^{-1}) ={\cal L} (l), \qquad  {\cal M} (l^{-1}) = {\cal M} (l) 
\end{equation} 
\begin{equation}\label{sym2}
{\cal H} (l^{-1}) ={\cal H} (l), \qquad  {\cal R} (l^{-1}) = {\cal R} (l),
 \qquad  {\mathbf \Lambda} (l^{-1}) = {\mathbf \Lambda} (l)
\end{equation} 

\medskip

The QH magnitudes are complete variables covering both classical and quantum, 
Planckian and super Planckian domains. They are more complete magnitudes than the $ Q$ or $H$ 
magnitudes alone which cover only one phase or domain: classical gravity or quantum/semiclassical domain. 

\bigskip

{\bf The complete (QH) de Sitter density.}
Let us complete the set of physical de Sitter magnitudes with the classical, 
quantum and QH de Sitter densities respectively, ($\rho_P$ being the Planck density scale):
($\rho_H$, $\rho_Q$, $\rho_{QH}$): 
\begin{equation}\label{rhoH}
\rho_H = \rho_P \left(\frac{H}{h_P}\right)^2 = \rho_P \left(\frac{\Lambda}{\lambda_P}\right), \qquad 
\rho_P = \frac{3 \; h_P^2}{8 \pi G}, \quad \lambda_P = \frac{3\; h_P^2}{c^4}
\end{equation} 
\begin{equation}\label{rhoQ}
\rho_{Q} = \rho_P \left(\frac{H_Q}{h_P}\right)^2 = \rho_P \frac{\Lambda_Q}{\lambda_P} = \frac{\rho_P^2}{\rho_H}
= \rho_P \left(\frac{h_P}{H}\right)^2 = \rho_P \left(\frac{\lambda_P}{\Lambda}\right)
\end{equation}                             
\begin{equation}\label{rhoQH}
\rho_{HQ} = \rho_{H} + \rho_{Q} = \rho_P \left(\frac{H_{HQ}}{h_P}\right)^2 = \rho_P \frac{\Lambda_{HQ}}{\lambda_P} 
\end{equation}
From Eqs. (\ref{QH2}), (\ref{rhoQ}) it follows that:\begin{equation}
\rho_{HQ}= \frac{\rho_H}{[\; 1 + \rho_H / \rho_P\; ]^2} 
= \frac{\rho_Q}{[\; 1 + \rho_Q / \rho_P \; ]^2}, 
\end{equation}
which satisfies 
$$
\rho_{HQ}\; (\rho_H)  =  \rho_{HQ}\; (\rho_Q) = \rho_{HQ} \;(\rho_P^2 / \rho_H),
$$
For small and high densities with respect to the Planck density $\rho_P$, the QH de Sitter density $\rho_{QH}$ behaves as: \begin{equation} \label{RhoQH1} \rho_{QH} \;(\rho_H << \rho_P) =  \rho_H \;[\; 1 - 2 (\rho_H / \rho_P) \;] 
+ O\;(\rho_H / \rho_P)^2  
\end{equation}
\begin{equation} \label{RhoQHP}
\rho_{QH} \; (\rho_H = \rho_Q = \rho_P)  = \frac{1}{4}\rho_P : \;\;\mbox {(Planck regime)}
\end{equation}
\begin{equation}  \label{RhoQH2}
\rho_{QH} \;(\rho_H >> \rho_P) =   \rho_Q \;[\; 1 - 2 (\rho_Q / \rho_P) \;] + O\;(\rho_Q / \rho_P)^2, 
\end{equation}

corresponding to the classical/semiclassical de Sitter regime (and its quantum corrections) 
Eq.(\ref{RhoQH1}), to the Planck scale de Sitter state Eq.(\ref{RhoQHP}),
and to the highly quantum, super Planckian, de Sitter phase Eq.(\ref{RhoQH2}). In the very 
classical regime, 
$\rho_{QH}$ is proportional to the classical density $\rho_H$, as it must be. In the highly 
quantum regime, $\rho_{QH}$ is proportional to $\rho_Q$, as it must be too.

\section {Classical and Quantum Dual de Sitter Temperatures and Entropies}

We complete now the set of relevant intrinsic de Sitter magnitudes, by including the Temperature 
and the Hubble horizon area. The temperature $T_H $ of the classical de Sitter Universe and the temperature $T_Q$ of 
the quantum de Sitter Universe are consistently defined as ($\kappa _B$ is the Boltzmann
 constant) :
\begin{equation} \label{TH}
T_H = \frac{M_H c^2}{2 \pi \kappa _B} , \qquad  T_Q = \frac{M_Q c^2}{2 \pi \kappa _B}, 
\end{equation}

which from Eqs. (\ref{LH}),(\ref{MH}),(\ref{MQ}) yield:
\begin{equation}\label{TH2}
T_H = t_P \left(\frac{M_H}{m_P} \right) = t_P \left(\frac{L_H}{l_P} \right) = t_P \left(\frac{h_P}{H} \right)
= t_P \sqrt{\frac{\lambda_P}{\Lambda}}
\end{equation}
\begin{equation}\label{TQ2} 
T_Q = t_P \left(\frac{m_P}{M_H} \right) = t_P \left(\frac{l_P}{L_H} \right) = t_P \left(\frac{H}{h_P} \right)
= t_P \sqrt{\frac{\Lambda}{\lambda_P}}
\end{equation}
$t_P$ being the Planck temperature, $T_Q$ and $T_H$ satisfy:
\begin{equation}
T_Q = \frac{t_P^2} {T_H}, \qquad t_P = \frac{m_P c^2}{2\pi \kappa _B} 
\end{equation}

\medskip

We see that the Quantum de Sitter Temperature $T_Q$ is the Hawking-Gibbons de Sitter temperature 
\cite{GibbHawkTemp}. 
This is the Quantum dual of the Classical de Sitter temperature $T_H$.

\medskip

The classical and quantum de Sitter areas $ A_H, A_Q$ are defined as:
\begin{equation} \label{AH}
A_H = 4 \pi L_H^2, \qquad   A_Q = 4 \pi L_Q^2,  
\end{equation}

which, from Eqs. (\ref{LH}),(\ref{MH}), (\ref{MQ}), (\ref{TH2}),(\ref{TQ2}) yield:
\begin{equation} \label{AHL}
A_H = a_P\left(\frac{L_H}{l_P}\right)^2 = a_P\left(\frac{M_H}{m_P}\right)^2 
= a_P\left(\frac{T_H}{t_P}\right)^2 = a_P\left(\frac{h_P}{H}\right)^2        
\end{equation}
\begin{equation} \label{AQL}
A_Q = a_P\left(\frac{l_P}{L_H}\right)^2 = a_P\left(\frac{m_P}{M_H}\right)^2 = 
a_P\left(\frac{t_P}{T_H}\right)^2 = a_P\left(\frac{H}{h_P}\right)^2
\end{equation}
$a_P$ being the Planck area. $A_Q$ and $A_H$ satisfy
\begin{equation}
A_Q = \frac{a_P^2}{A_H}, \qquad  a_P = 4 \pi\; l_P^2
\end{equation}

The classical and quantum areas are dual to each other through the Planck scale area $a_P$, 
and have the expressions:
\begin{equation}\label{AMTQ}
\frac{A_H}{a_P} =  \frac{a_P}{A_Q} = \frac{T_H}{T_Q} = \frac{1}{2\pi \kappa_B}\frac{M_H\;c^2}{T_Q}
\end{equation}
\begin{equation}\label{AMTH}
\frac{A_Q}{a_P} =  \frac{a_P}{A_H} = \frac{T_Q}{T_H} = \frac{1}{2\pi \kappa_B}\frac{M_Q\;c^2}{T_H},
\end{equation}

\medskip

which are entirely {\it symmetric} under the change $H\leftrightarrow Q$. 
Interestingly enough, the areas can be expressed as (one half) the ratio of 
the energy over the temperature, Eqs.(\ref{AMTQ}),(\ref{AMTH}), which is a typical 
{\it entropy} expression.
 The corresponding gravitational entropies $S_H$, $S_Q$ are:
\begin{equation} \label{SH}
S_H = \frac{\kappa_B}{4} \;\frac{A_H}{l_P^2}, \qquad S_Q = \frac{\kappa_B}{4}\;\frac{A_Q}{l_P^2} 
\end{equation}

Eq.(\ref{AMTQ}) is the Gibbons-Hawking or classical/semiclassical gravity 
de Sitter entropy. Eq.(\ref{AMTH}) is its quantum dual gravity de Sitter entropy.
From Eqs.(\ref{AMTQ}),(\ref{AMTH}):
\begin{equation} \label{SQ}
S_Q  = \frac{s_P^2}{S_H} , \qquad s_P =  \frac{\kappa_B}{4}\;\frac{a_P}{l_P^2} = \pi \kappa_B  
\end{equation}
$s_P$ being the Planck entropy. $S_H$ and $ S_Q$ read:
\begin{equation} \label{SHL}
S_H  = s_P \left(\frac{L_H}{l_P}\right)^2 = s_P \left(\frac{h_P}{H}\right)^2 = 
s_P \left(\frac{\lambda_P}{\Lambda}\right)  
\end{equation}     
\begin{equation} \label{SQL}
S_Q  = s_P \left(\frac{l_P}{L_H}\right)^2 =  s_P \left(\frac{H}{h_P}\right)^2 
= s_P \left(\frac{\Lambda}{\lambda_P}\right)
\end{equation}

\medskip

The classical and quantum entropies $S_H, S_Q$ satisfy the classical-quantum 
duality through the Planck scale entropy $s_P$, and have the expression:
\begin{equation} \label{SHMT}
\frac{S_H}{s_P} =\frac{1}{4 \pi \kappa_B}\;\frac{M_H}{T_Q} c^2, \;\;
\qquad
\frac{S_Q}{s_P} =\frac{1}{4 \pi \kappa_B}\;\frac{M_Q}{T_H} c^2,
\end{equation}
which is a typical entropy expression in terms of Mass and Temperature.

\section {Total de Sitter Temperature and Entropy}

From the total (QH) de Sitter magnitudes above discussed as 
the whole QH radius $L_{QH}$ and associated mass $ M_{QH}$, 
we define the corresponding total de Sitter temperature 
$T_{QH}$, area $A_{QH}$ of the $QH$ Hubble radius and entropy $S_{QH}$.

\medskip

The $QH$ temperature is defined as:
\begin{equation} \label{TQH1}
T_{QH} = \frac{M_{QH}c^2}{2\pi \kappa_B} ,
\end{equation}
which from Eqs (\ref{MQH}),(\ref{TH2}),(\ref{TQ2}), read
\begin{equation}\label{TQH2}
T_{QH} = \frac{m_P c^2}{2\pi \kappa_B} \; (\frac{M_H}{m_P} + \frac{m_P}{M_H}) 
= t_P\;(\frac{T_H}{t_P} + \frac{t_P}{T_H}),  
\end{equation} 
\begin{equation} \label{TQH3}
T_{QH} = (T_H + T_Q)\;\; \mbox{invariant under $H \leftrightarrow Q$}
\end{equation}

Consistently, the QH de Sitter Temperature is the sum of the classical gravity
de Sitter temperature $T_H$ plus the quantum (or semiclassical) Gibbons-Hawking de Sitter 
temperature $T_Q$. In terms of $H_{QH}$, the QH temperature $T_{QH}$ simply reads:
\begin{equation} 
T_{QH} = t_P \left(\frac{h_P}{H_{QH}}\right), \;\;\mbox{or}\;\;\frac{T_{QH}}{t_P} \equiv {\cal T} 
= {\cal H}^{-1} 
\end{equation}
Explicitely:
\begin{equation} 
T_{QH} = t_P (\frac{h_P}{H})\; [\;1 + (\frac{H}{h_P})^2\;] =
t_P (\frac{H}{h_P})\; [\;1 + (\frac{h_P}{H})^2\;],
\end{equation}
which is entirely invariant under the interchange $H \leftrightarrow h_P$. 

The $QH$ Area of the Hubble horizon follows from the QH Hubble radius $L_{QH}^2$ 
Eq.(\ref{LQH}):
\begin{equation} \label{AQH}
A_{QH} =  4 \pi   L_{QH}^2 =
2 \; (4 \pi l_P^2)\;\left[1 + \frac{1}{2}\; [\;(\frac{L_H}{l_P})^2 + (\frac{l_P}{L_H})^2\;]\right],
\end{equation}

which from  Eqs.(\ref{AH}),(\ref{AHL}),(\ref{AQL}) expresses as:
\begin{equation} \label{AQH2}
A_{QH} =  2\; a_P\left[ 1 + \frac{1}{2}\; [\;\frac{A_H}{a_P} + \frac{a_P}{A_H}\;] \right],
\qquad a_P = 4 \pi l_P^2
\end{equation}

The $QH$ area $A_{QH}$ is thus the sum of the Planck area $a_P$, the classical area $A_H$ and the quantum area $ A_Q$:
\begin{equation} \label{AQH3} A_{QH} =  2 a_P +  A_H \;+ \;  A_Q \end{equation}
$$ A_{QH} (A_H = a_P = A_Q) = 4 \; a_P $$
In units of the Planck area $a_P$, each of the areas can be in turn expressed 
as a energy over the temperature ratio  Eqs.(\ref{AMTH}),(\ref{AMTQ}), which yields:
\begin{equation} \label{AQHMT}
\frac{A_{QH}}{a_P} = 2 \; \left[\; 1 + \frac{1}{4 \pi \kappa_B} [\; \frac{M_H c^2}{T_Q} + 
  \frac{M_Q c^2}{T_H}\; ]\;\right] 
 \end{equation}
The corresponding QH gravitational entropy $S_{QH}$ is given by
\begin{equation} \label{SQH5}
S_{QH} = \frac{\kappa_B}{4} \frac{A_{QH}}{l_P^2} =
 2 \kappa_B\; \left[\; \frac{a_P}{4 l_P^2}  + \frac{1}{2}\;(\; \frac{A_H}{4 l_P^2}  
+ \frac{A_Q}{4 l_P^2} \;)\; \right]
\end{equation}
Thus,
\begin{equation}\label{SQH6}
S_{QH} =  2\;[\; s_P + \frac{1}{2}\;( S_H + S_Q )\;], 
\qquad s_P = \frac{\kappa_B}{4} \frac{a_P}{l_P^2} = \pi \kappa_B
\end{equation}
$$ 
S_{QH}\;(S_H = s_P = S_Q) = 4 s_P = 4 \pi \kappa_B 
$$

The whole entropy $S_{QH}$ turns out to be the sum of the Planck entropy $s_P$, 
the classical entropy $S_H$ and the quantum entropy $S_Q$ Eqs.(\ref{SH}),(\ref{SQ}),(\ref{SHMT}),
 and have the expression:

\begin{equation}
\frac{S_{QH}}{s_P} = 2 \;\left[\; 1 + \frac{1}{4 \pi \kappa_B} \; 
(\;\frac{1}{8} \frac{M_H c^2}{T_Q} + \frac{1}{8} \frac{M_Q c^2 }{T_H}\;)\;\right] 
\end{equation}
\begin{itemize}
\item{We see how the concept of classical-quantum duality and the QH variables naturally accompass, unify and simplify the relationship between the classical and quantum gravity magnitudes and regimes, in particular this is well appropriated to discuss the gravitational temperature and entropy in the different, classical and quantum, gravity regimes.}
\item{
The concept of gravitational entropy is the same for any of the gravity regimes: classical, quantum, Planck scale and  quantum gravity or super-Planckian regimes:  $S_H, S_Q, s_P$ or $S_{QH}$,  namely: $Area /4 l_P^2$ in units of $\kappa_B$.}  
\item{
For a classical size, ie a large macroscopic gravitational object, or our universe of radius $L_H$, this is the classical/semiclassical area $A_H$ and so the classical/semiclassical gravitational entropy $S_H$, which is the known Gibbons-Hawking de Sitter entropy \cite{GibbHawkEntropy}.} 
\item{
For a quantum size, ie a quantum microscopic object or quantum universe, ie of size equal to the Compton length $L_Q$, this is the quantum dual area $A_Q$ and so the quantum dual entropy $S_Q$. For a Planck length object or universe this is the Planck 
entropy $s_P$.  The whole or complete $QH$ entropy $S_{QH}$ turns to be the sum of the three components, as it must be.}
\end{itemize}
In dimensionless variables:
\begin{equation}
{\cal T} \equiv \frac{T_{QH}} {t_P},  \qquad {\cal A } \equiv  \frac{A_{QH}} {a_P},  
\qquad {\cal S} \equiv \frac{S_{QH}} {s_P},
\end{equation}
\begin{equation}
t \equiv \frac{T_H}{t_P},  \qquad a \equiv  \frac{A_H}{a_P},  
\qquad s \equiv \frac{S_H}{s_P},
\end{equation}

Eqs.(\ref{TQH1}),(\ref{AQH2}),(\ref{SQH5}), simply read:
\begin{equation}
{\cal T} =  (t + \frac{1}{t}), \qquad {\cal A } =  
2 \;[\;1 + \frac{1}{2} \;(a + \frac{1}{a})\;], \qquad
\qquad {\cal S} = 2 \; [\; 1 + \frac{1}{2} \;(s + \frac{1}{s})\;]
\end{equation}
$$ {\cal T}(t = 1) = 2,\qquad {\cal A} (a = 1) = 4, \qquad {\cal S} (s = 1) = 4 $$ 

And their duality symmetry simply stands:
\begin{equation}
{\cal T}(t) = {\cal T}(t^{-1}), \qquad {\cal A} (a)  =  {\cal A} (a^{-1}),
\qquad {\cal S} (s) = {\cal S} (s^{-1}),
\end{equation}
which show the simplification in terms of the Planck units, natural to the problem.

\section{Classical, Semiclassical, Planckian and Super-Planckian de Sitter Regimes}

The complete QH radius $L_{QH} = L_{QH} (L_H, L_Q) = L_{QH} (L_H, l_P)$ and their corresponding 
 QH Hubble constant $H_{QH}$, $QH$ mass $M_{QH}$, and their constant Planck scale values 
$(l_P, h_P, m_P)$ only depending on $(c, \hbar, G)$, allow to characterize in a precise
way the classical, semiclassical, Planckian and quantum (super-Planckian) de Sitter regimes:

\begin{itemize}
\item{$L_{QH} = L_{QH}(L_H, L_Q) \equiv L_{QH} (H, \hbar) $ 
yields the {\it whole} (classical/semiclassical, Planck scale and quantum (super-Planckian) 
de Sitter universe.}
\item{$L_{QH} = L_H = L_Q$ yields the Planckian de Sitter state, 
(Planck length de Sitter radius, Planckian vacuum density and Planckian scalar curvature):
\begin{equation}
L_H = l_P,  \quad H = h_P, \quad \lambda_P = 3\; h_P^2, \quad R = r_P = 4 \;\lambda_P, 
\quad l_P = \sqrt{(\hbar G / c^3)}
\end{equation}  }
\item{$L_{QH} = L_H >> L_Q $,  ie $ L_H >> l_P $ , $ H << h_P $,
 yields the classical de Sitter space-time.}
\item{$L_{QH} = L_Q  >> L_H $, ie $ L_H << l_P $,  $H >> h_P$, 
(high curvature $R >> r_P = 4 \Lambda_P,$) \\ 
yields a full quantum gravity super Planckian (inside the Planck domain $0 < L_H \leq l_P$) de Sitter phase.}
\item{$L_{QH} >> L_Q$ ie $L_{QH} \rightarrow \infty$ for $ L_H \rightarrow \infty $, ie 
$H \rightarrow 0 $ ie $\Lambda \rightarrow 0$, 
(zero curvature) yields consistently the classical Minkowski space-time, equivalent to the limit  
$L_Q \rightarrow 0 $ ie $ l_P \rightarrow 0$  ($\hbar \rightarrow 0$).}
\end{itemize}
The three de Sitter regimes are characterized in a complete and precise way:
\begin{itemize}
\item{(i) {\it Classical and Semiclassical de Sitter Regimes}:
(Inflation and more generally the whole known 
-classical and semiclassical- Universe is within this regime):
$ l_p < L_H < \infty$, \; ie $0 < L_Q  < l_P$, $\;$  $0 < H < h_P$,$\;$ $ m_P < M_H < \infty$.} 
\item{ (ii) {\it Planck Scale de Sitter state with Planck curvature and Planck radius:}
$L_H = l_P , \;\; L_Q = l_P,\;\; H = h_P = c/l_P,\;\; M_H = m_P$.}
\item{(iii)  {\it Quantum Planckian and super-Planckian Regimes:}
$ 0 < L_H \leq l_P $,\; ie $ \infty < L_Q  \leq l_P $, \;\;$ h_P \leq H < \infty $,\;\; $ 0 < M_H < m_P$.} 
\end{itemize}
The two above classical and quantum de Sitter regimes  (i) and (iii)  are  duals of each other 
in the precise meaning of the classical-quantum or wave-particle duality through the Planck 
scale de Sitter state (ii). This is the Planck scale duality or 
 classical-quantum gravity duality Eqs.(\ref{O}),(\ref{OQG}),(\ref{Oop}) at work.

\section{Numbers and Cosmological Implications}

Let us see now some illustrative cosmological values for the relevant classical, quantum and 
Planck scale magnitudes above refered. Typical cosmological values are: \begin{itemize}
\item{{\bf For the Universe Today}:
$$ H = 100 \; h \;\frac{Km}{sec\; Mpc}, \quad h = 0.7,\quad
\rho_{crit} = 2.77 \; 10^{11} h^2 \; \frac{M_{sun}}{(Mpc)^3},\quad 
\rho_{\Lambda} = 0.7 \; \rho_{crit}$$}
\end{itemize}
{\bf Classical, Planck and Quantum Dual values of the Hubble Radius, Mass and Age of the Universe Today, 
are typically:}
$$ L_H = 1.2 \; 10^{28} cm = 1.2\; 10^{61}\;l_P,\quad l_P = 10^{-33} cm ,\quad
L_Q = 0.8 \; 10^{-61}\;l_P$$
$$ M_H = 1.5 \; 10^{48} gr = 1.5 \; 10^{53} m_P,\quad
m_P = 10^{-5} gr,\quad  M_Q = 0.67 \; 10^{-53} m_P$$
$$T_H = 0.4 \; 10^{18} sec = 4 \; 10^{61}\; t_P,\quad
t_P = 10^{-44} sec,\quad
T_Q = 0.2 \;10^{-61} t_P$$
{\bf Classical, Planck and Quantum Dual values of the Hubble Constant, Cosmological Constant and Density of the Universe Today are typically:}
$$ H = 2.5 \; 10^{-17} sec^{-1} = 2.5\; 10^{-61} h_P,\quad
h_P = 10^{44} sec^{-1},\quad
H_Q = 10^{61} h_P$$
$$\Lambda = 3\; 10^{-34} sec^{-2} = 10^{-122} \lambda_P,\quad
\lambda_P = 3 \; 10^{88} sec^{-2},\quad
\Lambda_Q = 10^{122} \lambda_P$$
$$\rho_H = 10^{-29} \frac{gr}{cm^3} = 10^{-122} \rho_P,\quad
\rho_P = 10^{93}\frac{gr}{cm^3},\quad
\rho_Q = 10^{122} \rho_P$$
The values above correspond to the classical Universe today (subscript H), the 
Planck values (subscript P) and the Quantum dual values (subscript Q).
\begin{itemize}
\item{{\bf For the CMB era, Classical and Quantum values of the Age, Hubble constant,           
Size and Density of the Universe are typically:}}
\end{itemize}
$$L_H = 10^{24}\;cm =  10^{57} l_P,\quad
T_H = 10^{13} sec = 10^{57} t_P,\quad H = 10^{-57} h_P,\quad \rho_H = 10^{-114}\rho_P$$
$$L_Q = 10^{-57}l_P,\quad 
T_Q = 10^{-57}\; t_P,\quad H_Q = 10^{57} h_P,\quad  \rho_Q = 10^{114}\rho_P
$$
\begin{itemize}
\item{{\bf For the Inflation era, Classical/semiclassical and Quantum dual 
values of the Hubble Constant, Horizon size and Inflaton Mass are typically:}}
\end{itemize}
$$L_H = 10^{-27} \;cm =  10^{6} l_P,\quad
T_H = 10^{6} t_P,\quad H = 10^{-6} h_P,\quad
M_H =  10^{6} m_P$$
$$L_Q = 10^{-6}l_P,\quad
T_Q = 10^{-6}\; t_P,\quad H_Q = 10^{6} h_P,\quad  
\quad M_Q = 10^{-6} m_P, 
 $$
\begin{itemize}
\item{{\bf For the Solar system}: $M_{sun} = 10^{33} gr = 10^{38} m_P, 
\quad M_{Q\; sun} = 10^{-38}\; m_P$                    
$$M_{moon} = 7\; 10^{25} gr = 7 \; 10^{30}\; m_P,\; 
\qquad M_{Q\; moon} = 0.14\; 10^{-30} \;m_P$$
$$M_{asteroid,\;comet} = 10^{15} gr =  10^{20}\; m_P,\; 
\qquad M_{Q\; asteroid,\;comet} =  10^{-20}\;m_P$$}
\item{{\bf For Human scales}: $M_{human} = 10^{5} gr = 10^{10} \;m_P, 
\quad M_{Q\; human} = 10^{-15} gr = 10^{-10}\; m_P$                    
$$L_{human} = 1.7\; 10^{2} cm = 1.7 \; 10^{35}\; l_P,\; 
\qquad L_{Q\; human} = 10^{-68} cm = 10^{-35}\;l_P$$}
\item{{\bf For atomic scales}: $L_{atom} = 10^{20} \;l_P, \; \qquad  
T_{atom} = 10^{20} \;t_P, \; \qquad   M_{atom} = 10^{-20} \;m_P$   
$$L_{Q\;atom} = 10^{-20}\;l_P, \; \qquad  T_{Q\;atom} = 10^{-20}\; t_P, \; 
\qquad   M_{Q\;atom} = 10^{20}\; m_P $$}
\item{{\bf For elementary particles (ex.the electron mass)}: 
$M(eV/c^2) = 10^{-33} gr = 10^{-28}\; m_P,\quad M_Q(eV/c^2) = 10^{23} gr = 10^{28}\;m_P$}

\item{We see that the elementary particle masses do appear as the {\it quantum duals through 
the Planck scale} of  the typical solar system objects. For instance, the quantum dual of a typical comet or asteroid 
mass say is a  typical atomic mass.  The quantum dual of the electron mass 
$M_Q(eV/c^2)$ is a typical moon mass of $10^{22}$ kgr.}
\item{That is to say, there is a {\it physical classical-quantum duality} through the Planck mass or correspondence between the {\it macroscopic or astronomical gravitational 
masses/sizes} and the {\it elementary particle and quantum masses/sizes}, the Planck scale being 
the {\it crossing or inversion scale} of the two mass/size domains: ${M_Q = {m_P^2}/{M}}$. 
These two domains are {\it precisely} connected through the  classical-quantum 
(ie wave-particle, Compton, de Broglie) duality 
including gravity Eqs. (\ref{O}),(\ref{MH}),(\ref{MQ}): namely, a Planck scale duality 
or classical-quantum gravity duality Eqs.(\ref{O})-(\ref{Oop}).}
\item{
Notice that in the {\bf classical CMB era} (at about $3.8 \; 10^5 yr = 10^{13} sec$), 
the gravitational size $L_H$, 
age and temperature $T_H$ of the Universe are, as reported above, equal to $10^{57}$ 
(in Planck units), while their quantum duals in the quantum precursor era are 
$10^{-57}$.  The classical gravitational entropy $S_H$ and the $H$-associated density 
$\rho_H$ in the classical CMB era are respectively:
$$S_H = 10^{114} s_P, \;\qquad \rho_H = 10^{-114},\; \qquad s_P = \pi \kappa_B$$ 
Their quantum dual values in the quantum CMB precursor era $10^{-57} t_P$
being respectively $S_Q = 10^{-144} s_P$ and
$\rho_Q = 10^{144} \rho_H$.   
The quantum entropy $S_{Q}$ in the quantum CMB precursor era is extremely low as it must be, 
because it will increase along the Universe evolves and classicalizes, reaching its 
classical gravitational value $S_{H} = 10^{114}$ in the classical known CMB era (arrow of time).}
\item{Let us recall the classical entropy $S_{cmb}$ of the CMB black body radiation contained in the classical
Hubble volume: $S_{cmb} = (4/3) \pi s_{\gamma} H^{-3}$,  
the total number of the CMB photons being $1.5 \; 10^{89}$, the average entropy 
per photon $3.6 \;\kappa_B$, hence:$$ S_{cmb} =  1.72 \;10^{89} s_P, \; \qquad 
S_{Q\; cmb} =  \frac{s_P^2}{S_{cmb}} = 0.58 \; 10^{-89} s_P, $$
The gravitational entropy $S_H$ is the dominant component in the classical CMB era,  
and represents un {\it upper bound} for the CMB photon radiation entropy $S_{cmb}$ in it. 
The quantum gravitation entropy $S_Q$ is the smaller component and is a lower bound for the 
$S_{Q\; cmb}$ value in the quantum precursor era. Similarly, for the respective 
Temperatures: The classical and quantum dual gravitational 
temperatures $T_H$ and $T_Q$ at the CMB age are
$$ T_H = 10^{57} t_P, \qquad T_Q = 10^{-57} t_P, 
\qquad t_P = 10^{32}\; K  $$
The classical and quantum dual Temperatures of the CMB radiation are:
$$ T_{cmb} = 2.73 K = 2.73 \; 10^{-32} \; t_{P}, \qquad
T_{Q\;cmb} = 0.37 \; 10^{32} \;t_{P}$$
The gravitational temperature $T_H$ in the classical CMB era,
represents un {\it upper bound} for the CMB photon radiation Temperature $T_{cmb}$ in it. 
The quantum gravitation temperature $T_Q$ is the smaller component and is a lower bound for the $T_{Q\; cmb}$ value in the precursor era.}
\end{itemize}

Relevant implications for Inflation and Dark Energy, are discussed in detail in Sections IX and X below.

\bigskip

{\bf Quantum Inflationary Fluctuations, the Gibbons-Hawking Temperature and CMB anisotropies}:
Interestingly enough, the power spectra of quantum primordial fluctuations 
of Inflation can be expressed in terms of the Quantum
and Planck temperatures $T_Q$, $t_P$. As is known, the spectra of inflationary scalar 
curvature and tensor perturbations are given by:
\begin{equation}
\Delta_{k, H}^S = \frac{1}{\sqrt{\pi \epsilon}} \; \frac{H}{m_P}, 
\qquad
\Delta_{k, H}^T = \frac{4}{\sqrt{\pi}}\;\frac{H}{m_P}, 
\end{equation}
$\epsilon$ being the slow-roll parameter. From Eq.(\ref{TQ2}) they can be expressed in terms 
of $ T_Q/t_P$ as:
\begin{equation}
\Delta_{k, H}^S = \frac{1}{\sqrt{\pi \epsilon}} \; \frac{T_Q}{t_P}, 
\qquad
\Delta_{k,H}^T = \frac{4}{\sqrt{\pi}}\;\frac{T_Q}{t_P} 
\end{equation}
Thus:
\begin{equation}
 T_Q = t_P \; \sqrt{\pi \epsilon}\; \Delta_{k, H}^S = t_P \; \frac{\sqrt{\pi}}{4} \; \Delta_{k, H}^T 
\end{equation}
Or, in terms of the ratio $r$: 
$$ 
T_Q= t_P \; \frac{\sqrt{\pi r}}{4} \Delta_{k, H}^S,   \qquad
r = \frac{[\Delta_{k, H}^T]^2}{[\Delta_{k, H}^S]^2}
$$
Therefore, for the amplitude value $\Delta_{k, H}^S$ from the CMB data \cite{WMAP1},\cite{Planck6},
we get  for $ T_Q$:
$$T_Q = \sqrt{\pi \epsilon} \; 10^{28} \; K = \sqrt{\pi \epsilon} \; 10^{-4} \; t_P 
\; \qquad \mbox{and} \;\qquad T_Q = \frac{\sqrt{\pi\;r}}{4} \; 10^{-4} \; t_P$$
From the last recent bound  $r < 0.07$ \cite{Planck6}:
\begin{equation}
 T_Q <  1.169 \; 10^{-5}\; t_P =  1.169 \; 10^{27}\; K
\end{equation}
We see that $T_Q$ for classical/semiclassical Inflation is constrained to be less than $10^{-5} t_P$, 
{\it consistent} with the semiclassical gravity character of Inflation.
Interestingly,   
the Quantum Temperature $T_Q$, which is here precisely the {\it Hawking-Gibbons de Sitter temperature},  
can be measured or constrained through the real CMB data which constrain 
Inflation. This is important because:
{\bf (a)} The conceptual quantum/semiclassical gravity 
nature of the Hawking-Gibbons de Sitter temperature, 
and {\bf (b)} Contrary to the Inflation case, the  equivalent Hawking temperature for astrophysical black holes 
cannot be experimentally measured since it is extremely low: lower than the CMB temperature. 
$T_Q$ could be higher for small or primordial black holes but these have not been  
 detected. 

\bigskip

\section {Implications for Inflation}

We discuss here in more detail the consequences of the Q and QH observables for Inflation.
Recall that in Classical Inflation, at first order in the slow-roll expansion, 
the scalar curvature and tensor perturbation spectra are given by:
\begin{equation}\label{DeltaSO}
[\Delta^S_{k,\;H}]^2 =  \frac{1}{\pi \epsilon}\;\left(\frac{H}{m_P}\right)^2,
\qquad
[\Delta^T_{k, \;H}]^2 =  \frac{16}{\pi }\;\left(\frac{H}{m_P}\right)^2
\end{equation}
($\Delta^S_{k, \;H}, \Delta^T_{k, \;H}$ stand here for the scalar and tensor fluctuations respectively).
The slow roll parameters are given by:
\begin{equation}\label{epsilon}
\epsilon = \frac{m_P^2}{4\pi} \left(\frac{H'}{H}\right)^2, \qquad
\eta = \frac{m_P^2}{4\pi} \left(\frac{H''}{H}\right), 
\quad \xi = \frac{m_P^2}{2\pi} \left(\frac{H'\;H'''}{H^2}\right)
\end{equation}
H' and H'' stand for the first and second derivatives with respect to the inflaton field. 
$\epsilon \approx \eta << 1 $ are first order in slow roll, $\xi$ 
second order slow roll parameter,   
with the hierarchy $\xi = O (\epsilon^2)$, and following so on in the slow roll expansion.

The slow roll parameters are related to the observables ratio $r$ and spectral scalar index $n_s$
by: \begin{equation}\label{parameters}
\epsilon = \frac{r}{16},\qquad \eta = \frac{1}{2}(\;n_s - 1 + \frac{3}{8}\;r\;),
\qquad           
\xi = \frac{r}{4}(\;n_s - 1 + \frac{3}{16}\;r - \frac{1}{2}\frac{dn_s}{dlnk}\;)            
\end{equation}
The difference $(\epsilon - \eta)$ is a measure of the {\it departure from scale invariance} at first order in slow roll:
\begin{equation} \label{Delta}
\Delta \equiv (\epsilon - \eta) = \frac{1}{2}(\;n_s - 1\;) + \frac{r}{8}
\end{equation}
From Eqs. (\ref{DeltaSO}),(\ref{QH2}) the Quantum (QH) generalization of the power spectrum of scalar curvature and tensor perturbations is given by:
\begin{equation}\label{DeltaSQH}
[\Delta^S_{k, \;QH}]^2 =  \frac{1}{\pi \epsilon_{QH}}\;\left(\frac{H_{QH}}{m_P}\right)^2
= \frac{1}{\pi \epsilon_{QH}}\;\left(\frac{H}{m_P\;[\; 1 + (H/h_P)^2 \;] }\right)^2
\end{equation}
\begin{equation} \label{DeltaTQH}
[\Delta^T_{k, \;QH}]^2 =  \frac{16}{\pi }\;\left(\frac{H_{QH}}{m_P}\right)^2
= \frac{16}{\pi}\;\left(\frac{H}{m_P\;[\; 1 + (H/h_P)^2 \;]}\right)^2
\end{equation}

\medskip

Here $\epsilon_{QH} << 1$ is the first order QH slow roll parameter, we compute it below,
and $ h_P$ is the Planck Hubble constant value. Thus, the QH inflationary spectra get expressed as:
\begin{equation}
[\;\Delta^S_{k, \;QH}\;]^2 = [\;\Delta^S_{k, \;H}\;]^2\; \left(\frac{1}{[ \;1 + (H/h_P)^2 \;]^2 }\right)\; 
\frac{1}{( 1 - \delta \epsilon_{QH})}
\end{equation}
\begin{equation}
[\;\Delta^T_{k, \;QH}\;]^2 = [\;\Delta^T_{k, \;H}\;]^2 \; \left(\frac{1}{[\; 1 + (H/h_P)^2 \;]^2 }\right)
\end{equation}

\medskip

where $[\Delta^S_{k,\;H}]^2$ and  $[\Delta^T_{k, \;H}]^2$ are the standard spectra of scalar 
curvature and tensor perturbations in Classical H Inflation Eqs.(\ref{DeltaSO}). 

\medskip

Each QH power spectrum  $[\Delta^{S,T}_{k, \;QH}]$ is expressed in terms of each Classical H Inflation  spectrum 
$[\Delta^{S,T}_{k, \;H}]$ and it is 
modified by a factor $[\; 1 + (H /h_P)^2\;]^{-2}$ arising from $H_{HQ}$.  
In addition, $[\Delta^S_{k,\;QH}]^2$ gets also modified by the QH factor $(1 - \delta \epsilon_{QH})^{-1}$ arising from the QH slow parameter $\epsilon_{QH}$:
\begin{equation}
\epsilon_{QH} = \frac{m_P^2}{4\pi} \left(\frac{H'_{QH}}{H_{QH}}\right)^2 \equiv \epsilon \;(\; 1 - \delta \epsilon_{QH}\;),
\end{equation}
where $\epsilon$ is the standard slow-roll parameter Eq.(\ref{epsilon}) and $\delta \epsilon_{QH}$ is its QH modification given by:
\begin{equation} \label{deltaepV}
\delta \epsilon_{QH} = 2 \;\frac{H H_{QH}}{h_P^2}\;[\; 1 - \frac{H H_{HQ}}{2 h_P^2}\;] 
\;= \;
\frac{4 (H/h_P)^2}{[\; 1 + (H/h_P)^2\;]} \left[\; 1 - \frac{(H/h_P)^2}{[\; 1 + (H/h_P)^2\;]}\;\right]
\end{equation}

The ratio $r_{QH}$ turns out precisely modified by this $\delta \epsilon_{QH}$ factor:

\begin{equation}
r_{QH} = \frac{[\;\Delta^T_{k,\;QH}\;]^2}{[\;\Delta^S_{k,\;QH}\;]^2} = 
\frac{[\;\Delta^T_{k\;H} ]^2}{[\;\Delta^S_{k\; H}]^2}\; (\;1 - \delta \epsilon_{QH}\;)
\end{equation}
Thus,
$$r_{QH} =  r \;(1 - \delta \epsilon_{QH}),\qquad  r_{QH} = 16 \;\epsilon_{QH},
 \qquad r = 16 \;\epsilon $$
with $\delta \epsilon_{QH}$ given by Eq.(\ref{deltaepV}).
The QH slow roll parameter $\eta_{QH}$ is given by:
\begin{equation}
\eta_{QH} = \frac{m_P^2}{4\pi} \left(\frac{H_{QH}''}{H_{QH}}\right),
\end{equation}
which from  Eq.(\ref{QH2}) can be recasted as:
\begin{equation}
\eta_{QH} = \left[\;\eta -\frac{1}{2} m_P \sqrt{\frac{\epsilon}{\pi}}\;\frac{H H_{QH}}{h_P^2}\; \right]
\left[\; 1- \frac{H H_{QH}}{h_P^2}\;\right]
\end{equation}
\begin{equation}
\eta_{QH} = \left[\;\eta -\frac{1}{2} m_P \sqrt{\frac{\epsilon}{\pi}}\;
\frac{(H/h_P)^2}{[\; 1 - (H / h_P)^2\;]}\;\right]
\left[\; 1- \frac{(H/h_P)^2}{ [\; 1 - (H / h_P)^2\;]}\;\right]
\end{equation}

\medskip

where $\epsilon$ is the standard slow roll parameter of classical H inflation Eq.(\ref{epsilon}). 
The QH modifications express themselves in terms of  $(H/h_P)^2 = (l_P H/c)^2$ and powers of it.

Is also of interest to compute the QH departure from scale invariance which is given by
\begin{equation} \label{DeltaQH}
\Delta_{QH} \equiv (\eta_{QH} - \epsilon_{QH})= \frac{1}{2}(\;n_{s\; QH} - 1 \;) + \frac{r_{QH}}{8}
\end{equation}

{\it Typically, for (classical)  Inflation}:  $H = 10^{-6} h_P$, ie $H << h_P$ and we can safely 
expand the above QH Inflation expressions in powers of $(H/h_P)^2$, namely:
$$\epsilon_{QH} (H << h_P) = \epsilon\; [\; 1 - 4\;(l_P H)^2 + 
O (l_P H)^4 \;]$$
$$\eta_{QH}\; (H << h_P) = \eta \; [\; 1 - 2\;(l_P H)^2 ] - 
m_P \sqrt{\frac{\epsilon}{\pi}}\;(l_P H)^2 + O (l_P H)^4 \;]$$
Thus, at first order in $(H/h_P)^2$:
\begin{equation}
[\;\Delta^S_{k, \;QH}\;]^2 = [\;\Delta^S_{k, \;H}\;]^2\; [\; 1 + 2\;(l_P H)^2 + O (l_P H)^4 \;]  
\end{equation}
\begin{equation}
[\;\Delta^T_{k, \;QH}\;]^2 = [\;\Delta^T_{k, \;H}\;]^2 \; [\; 1 - 2\;(l_P H)^2 + O (l_P H)^4 \;]  
\end{equation}
\begin{equation} \label{rQH2}
r_{QH} = r \;[\; 1 - 2\;(l_P H)^2 + O (l_P H)^4 \;]
\end{equation}

And the {\bf QH departure from scale invariance} Eq.(\ref{DeltaQH}) gets corrected as:
\begin{equation} \label{DeltaQH2}
\Delta _{QH} = \Delta\;[\; 1 - 2 \; (l_P H)^2 \;] + \sqrt{\epsilon} \;(l_P H)^2
\; [\;2 \sqrt{\epsilon} - \frac{m_P}{\sqrt{\pi}}\;] + O (l_P H)^4
\end{equation}
where $\epsilon$ is the standard slow roll parameter of classical H inflation 
Eqs.(\ref{epsilon})-(\ref{parameters}),  
and $\Delta$ is the {\it departure from scale invariance} of the classical H Inflation Eq.(\ref{Delta}) 
meaning that $(n_s - 1)$ {\it and} $r$ {\it are not zero}. In terms of the observables $(n_s, r)$ and $(n_{s\;QH}, r_{QH})$, Eq.(\ref{DeltaQH2}) yields for $n_{s\;QH}$:
 \begin{equation} \label{nsQH}
n_{s\;QH} =  n_s \;[\; 1 + 2 \;(l_P H)^2 \;] - 2 \;(l_P H)^2\; [\; 1 + m_P \sqrt{\frac
{\epsilon}{\pi}} \;] + O(l_P H)^4  
\end{equation}
where Eqs.(\ref{rQH2}),(\ref{DeltaQH}) and (\ref{Delta}) have been used.
Typically, for classical Inflation: $L_H = 10^{6}\;l_P$, 
the total QH corrections are thus: 
$$\frac{r_{QH}}{r} - 1 = - 2 \;10^{-12}$$
$$\frac{n_{s\;QH}}{n_s} - 1 =
2 \;10^{-12}\;[\;1 - \frac{1}{n_s} (\;1- \frac{m_P}{2}\sqrt{\frac{\epsilon}{\pi}}\;)\;]$$
\begin{itemize}
\item{We see that the complete QH quantities allow to get {\it quantum corrections to Inflation 
and its observables in a direct, simple and consistent way}}. 
\item{Notice the {\it sign} 
of the corrections: The quantum gravity QH corrections {\it enhance} the scalar curvature spectrum and {\it reduce} the tensor perturbations.
The QH corrections are of the same order of magnitude and {\it 
sign} as the {\it quantum inflaton} corrections computed in the Effective 
Theory of Inflation within the Ginsburg-Landau approach \cite{Boya2005}, \cite{Boya2006}, \cite{BDdVS}. 
This also shows the {\it robustness and reliability} of the slow roll approximation and the 
Effective Theory of inflation. [If the reduced Planck mass 
$M_P$ is used, $m_P = \sqrt{8\pi} \;M_P$: $(H/M_P)_{Inflation} = 10^{-5} = 10 \;(H/m_P)$ here].} 
\item{Notice that the QH factor modifying the Hubble constant and the inflationary spectra can be written as the summation of the series:
\begin{equation}\label{sum}
QH \equiv \frac{H}{[\;1 + (H / h_P)^2\;]} = H \; \sum _{n=0}^{\infty} (-1)^n \left(\frac{H}{h_P}\right)^2 
\end{equation} 
The QH factor covers the {\it full classical and quantum range}, namely:
If $H < h_P$, Eq.(\ref{sum}) yields the usual corrections in $(l_P H)^2$.
If $H >> h_P$, Eq.(\ref{sum}) precisely {\it changes to the quantum regime}, ie to the quantum Hubble rate 
$H_Q$, which is the {\it super-Planckian domain}:
\begin{equation}\label{sumQ}
HQ \equiv \frac{H_Q}{[\;1 + (H_Q / h_P)^2\;]}   
\end{equation}}
\item{In the case of classical Inflation: $H$ is about $10^{-6}\;h_P$, as we have seen.
In the case of the quantum precursor Inflation era at about $10^{-6}\;t_P$:
$H_Q$ is about $10^{6}\;h_P$,  
Whatever it be: in the classical phase 
 (after $t_P$) or in the quantum precursor phase (before $t_P$), Inflation occurs not too far from the Planck scale: $10^{\pm 6} t_P$ or $10^{\mp 6} h_P$.}
\end{itemize}

\section{Implications for Dark Energy}

Dark energy and its more direct candidate,
the cosmological constant, \cite{Riess},\cite{Perlmutter},\cite{Schmidt},\cite{WMAP1},\cite
{WMAP2},\cite{DES},\cite{Planck6} 
is relevant to both modern cosmology and particle physics. 
Let us recall the value of the observed dark energy density today $ \rho _H \equiv \rho_{\Lambda}$:
\begin{equation} \label{rhovalue}
\rho_\Lambda = \Omega_\Lambda \rho_c =  3. 28 \; 10^{-11} (eV)^4 = (2.39 \; meV)^4, \qquad  m eV= 10^{-3} eV
\end{equation}
corresponding to $ h = 0.73, \qquad \Omega_\Lambda = 0.76 , \qquad H = 1.558 \; 10^{-33} eV.$

\medskip

The last Planck satellite data yield the values \cite{Planck6}: 
\begin{equation} \label{Hvalue}
H = 67.4 \pm 0.5 \; Km \; sec^{-1}\; Mpc^{-1}, \quad \Omega_\Lambda h^2 = 0.0224 \pm 10^{-4}
\end{equation} and \begin{equation} \label{Omegavalue}
\Omega_\Lambda = 0.6847 \pm 0.0073, \quad  \Omega_\Lambda h^2 = 0.3107 \pm 0.0082,
\end{equation} 
which implies for the cosmological constant {\bf today}:
\begin{equation} \label{Lambdavalue} 
\Lambda = (4.24 \pm 0.11) \; 10^{-66}\; (eV)^2 = (2.846 \pm 0.076) \; 10^{-122}\; m_P^2
\end{equation} 
The density $\rho_\Lambda$ associated to $\Lambda$ Eq.(\ref{rhovalue}) 
is precisely:
\begin{equation} \label{rhoLambda} 
\rho_\Lambda = \frac{\Lambda}{8 \pi G} = \rho_P \left(\frac{\Lambda}{\lambda_P}\right),
\end{equation} 
where the Planck scale values $\rho_P,\lambda_P$ are:
$$\rho_P = \frac{\lambda_P}{8 \pi G} \quad \; \lambda_P = \ 3 h_P^2
$$
The enormous discrepancy between the large theoretical value expected from microscopic particle 
physics for the vacuum energy density $\approx 10^{122}$
and the small cosmological value observed today $\rho_{\Lambda}\approx 10^{-122}$ is largely known as the 
cosmological constant problem.
{\bf However, several clarifications are in order here}:

\medskip

{\bf(i) The classical gravity vacuum.} The $\Lambda$ value Eq. (\ref{rhovalue}), (\ref{Hvalue}), 
(\ref{Lambdavalue}) observed at the present era today corresponds to the {\it classical} 
(non quantum) value of the vacuum energy density of the {\it classical large Universe today}: 
large radius $L_\Lambda$ or large Age, large mass $M_\Lambda$ and large classical/semiclassical entropy $S_H$,
and thus small rate $H$, low temperature $T_\Lambda$, small $\Lambda$, {\it small and dilute classical 
vacuum density $\rho_{\Lambda}$} described here in the above sections. 

\medskip

{\bf(ii) The quantum gravity vacuum.} The $\Lambda$ density $\rho_{\Lambda}$ observed today is the {\it classical} vacuum density of the Universe today which is a gravitationally {\it classical, large} and 
{\it dilute} Universe. The value of $\rho_{\Lambda}$ and $\Lambda$
Eqs.(\ref{rhovalue}), (\ref{rhoLambda}) is precisely the {\it classical dual value} of 
the {\it quantum} cosmological constant value $\Lambda_Q$ Eq.(\ref{dualQH}), and therefore the classical dual value of the {\it quantum vacuum energy $\rho_Q$} Eq.(\ref{rhoQ}): This is precisely and clearly expressed in the following Eqs: 
\begin{equation} \label{LambdaHvalue} 
\Lambda = 3 H^2 = \lambda_P  \left (\frac{H}{h_P}\right)^2 = \lambda_P \left (\frac{l_P}{L_H}\right)^2
= (2.846 \pm 0.076) \; 10^{-122}\; m_P^2
\end{equation}
\begin{equation} \label{LambdaQvalue} 
\Lambda_Q = 3 H_Q^2 = \lambda_P \left (\frac{h_P}{H}\right)^2 = \lambda_P \left (\frac{L_H}{l_P}\right)^2
= (0.3516 \pm 0.094) \; 10^{122}\;m_P^2 
\end{equation}
\begin{equation} \label{LambdaQLambda} 
\Lambda_Q = \frac{\lambda_P^2}{\Lambda}, \qquad \lambda_P = 3 h_P^2 = 3 m_P^2 
\end{equation}
{\it The quantum dual value $\Lambda_Q$ is {\it precisely} the quantum vacuum value obtained 
from particle physics.}
\begin{equation} \label{rhoQii}
\rho_Q = \rho_P \left(\frac{\Lambda_Q}{\lambda_P}\right) 
= \frac{\rho_P^2}{\rho_\Lambda} = 10^{122}\; \rho_P
\end{equation}

{\bf(iii) The classical and quantum dual values}. That is to say, the two huge different values: $10^{-122}$ and $10^{122}$ 
(in Planck units) refer to 
{\it  two huge physically different} vacuum energies of the Universe corresponding to two huge different eras, to two huge different physical cosmological conditions (present time and very early eras), to two different vacuum states or regimes of the Universe, and consistently, they {\it must be different}. 
Such enormous difference must be in such way and is {\bf not} a problem or inconsistency:  Moreover and consistently, one value is the {\it quantum physics dual} of the other -or the quantum precursor of the other-  as expressed by Eqs.(\ref{LambdaHvalue}),(\ref{LambdaQvalue}),(\ref{LambdaQLambda}),(\ref{rhoQii}). 

\medskip

{\bf (iv)}. This is {\bf not fortuitous}, that is to say, this is not pure chance or unexplained coincidence.
{\bf (v)}. This is {\bf not trivial}, that is to say, this is simple, deep and robust.

\medskip

There is no problem between the two extremely different values $\Lambda$ 
and $\Lambda_Q$ or equivalently between $\rho_{\Lambda}$ and $\rho_Q$, because the two 
values {\it do not} refer to the same vacuum or eras: one is exactly the {\it 
classical} physics today vacuum energy density $\rho_{\Lambda}$, the other is  {\it its quantum 
dual} value in the planckian and superplanckian very early phase $10^{-61}\;t_P \leq t \leq t_P$: 
This early phase of the Universe is exactly the {\it quantum precursor} of the today classical era in the precise 
meaning of the wave-particle (or classical-quantum) duality including gravity,
Eqs.(\ref{LambdaHvalue}) to (\ref{rhoQii}). 

\medskip

The two different values are explained by the fact that they are exactly, mathematically and physically,
the classical-quantum dual of each other: {\it The $\Lambda_Q$ value 
Eq.(\ref{LambdaQvalue})-(\ref{LambdaQLambda}), that is to say, the vacuum value computed 
from particle physics 
is exactly the quantum dual value of the classical $\Lambda$ value observed today Eq.(\ref{LambdaHvalue}).}

\medskip

{\bf (vi) Crossing the Planck scale}. The two values: $\Lambda$ and $\Lambda_Q$, (or equivalently $\rho_
\Lambda$ and $\rho_Q$) 
refer to the same concept or nature of $\Lambda$ or $\rho_\Lambda$ as a vacuum energy 
density or cosmological constant but they are in two huge different vacuum states or two huge different cosmological 
epochs: Classical state and classical epoch today for $\Lambda$ observed today, and quantum state 
and quantum super-Planckian very early universe epoch for the quantum mechanical 
super-Planckian value $\Lambda_Q$.

\medskip

The classical value today $\Lambda = 3H^2$ corresponds to the classical Universe today 
of classical rate $H$  and classical cosmological radius $L_H = c/ H$. The quantum mechanical 
value $\Lambda_Q = 3 H_Q^2$ corresponds to the early quantum Universe of quantum rate $H_Q$ and
quantum radius $L_Q = l_P^2/L_H = \hbar/M_H c$ which is {\it exactly} the quantum dual of the classical
horizon radius $L_H$: $L_Q$ is {\it precisely} the quantum (Compton) length of the Universe for the
gravitational mass $M_H = L_H c^2/G$.

\bigskip

{\bf (vii) Two extremely different physical conditions and gravity regimes}. This is a realistic, clear and precise illustration of the {\it physical classical-quantum 
duality between the two extreme Universe scales and gravity regimes}: the dilute state and Horizon size 
of the Universe today on the one largest known side, and the super-Planckian scale and highest density state on the smallest side: Length, Mass, and their associated time (Hubble rate) and vacuum energy density 
($\Lambda, \rho_\Lambda$) of the Universe {\it today} are truly
{\it classical},  while its extreme past at $10^{-61}\; t_P$ =  
$ 10^{-105}$ sec deep inside the Planck domain of extremely small size and high vacuum density 
value ($\Lambda_Q, \rho_Q$) are truly {\it quantum and super-Planckian}.  

\medskip

This is the {\it classical-quantum or wave-particle duality} between  the classical macroscopic 
(cosmological) gravity physical domain and 
the quantum microscopic particle physics and super-Planckian domain 
through the {\it crossing} of the Planck scale, {\it Planck scale duality} in short.

\medskip

{\bf (viii) The true problem}. The huge difference between the two values $\Lambda = \rho_\Lambda = 10^{-122}$
and $\Lambda_Q = \rho_Q = 10^{122}$ is indeed {\it correct} and must
be such way, precisely because the two values refer to huge different physical conditions, regimes and 
states which are classical-quantum duals of each other. 

The two values refer to two different gravitational vaccua: classical, on one side (present time era), and full quantum super-Planckian energy on the opposite physical side (past remote era), 
and these are {\it two extreme different and dual energy density components}, $\rho_{\Lambda}$ and $\rho_Q$,  contributing 
to the {\it same total} vacuum energy of the Universe $\rho_{\Lambda Q}$. 

Namely, there is indeed a cosmological constant problem but the true problem is {\bf not}  
the huge discrepancy between the observed value today and the computed particle
physics value. The true problem is to know the origin and the nature (the type) 
of the predominant particle(s) 
associated to the vacuum energy density and how to identify and detect them.  

\medskip

{\bf (ix) A General framework}. This is not a tailored argument in order to explain solely one 
problem (dark energy) or one cosmological constant value. This is just one of the 
consequences or applications of a general clarifying simplifying framework which completes at the level 
of the classical and quantum observables, the classical/semiclassical gravity observables 
on the one hand, and the microscopic quantum particle physics, planckian and super-planckian 
magnitudes in the early quantum eras on the other hand, and connects them through the 
classical-quantum (wave-particle) duality, (and one of such observables is just the vacuum energy density). 

\medskip

 {\bf In Summary:} There is a {\it deep concept} behind the cosmological
vacuum energy density or cosmological constant: the classical-quantum (or wave-
particle) duality through the Planck scale, or Planck scale duality.
This extends to the Planckian and super-Planckian 
domain the classical-quantum duality of quantum theory and includes gravity in it: 
classical-quantum gravity duality or wave-particle-gravity duality.

Interestingly enough, including the Temperature and gravitationnal
Entropy of the Universe in the description {\it consistently} supports the 
cosmic classical-quantum duality  
and shed more insight in the cosmological constant/vacuum energy nature of the dark energy. 
We discuss it in the next section.

\section{The Cosmological Constant: 
Vacuum Energy, Entropy and Temperature of the Universe}

As we have seen, a key concept in order to understand the present value of the cosmological constant value 
and the so-called cosmological constant problem is the {\it classical-quantum duality}, 
precisely when applied to gravitational masses or objects, namely the classical-quantum duality 
through the Planck scale, or shortly Planck scale duality. The second important concept is the gravitational entropy 
$S_\Lambda$, namely the area 
of gravitational objects in units of $\kappa_B$ which is the Hawking-Gibbons de Sitter entropy, 
and its quantum dual entropy $S_Q$, as we will see below.

\medskip

The classical/semiclassical gravitational entropy $S_\Lambda$ of the Universe today is given by Eq.(5.13).
The Gibbons-Hawking de Sitter entropy is exactly $S_{\Lambda}$, 
while its quantum dual entropy $S_Q$ is given by Eq.(5.14).
The classical $\Lambda$-temperature $T_{\Lambda}$ of the Universe today is given by Eq.(5.2).
The quantum temperature $T_Q$ Eq.(\ref{TQ2}) is precisely the quantum dual temperature of $T_{\Lambda}$.
The Gibbons-Hawking de Sitter Temperature is exactly $T_Q$.

\medskip

Moreover, let us recall that the cosmic (de Sitter) gravitational entropy and temperature were first derived in the context of the euclidean (imaginary time) quantum gravity \cite{GibbHawkEntropy}: That is to say, the 
Wick rotated path integral or partition function of 
gravitation and matter fields which in the saddle point approximation yields the classical action
as the gravitational (Bekenstein-Hawking-Gibbons) entropy \cite{GibbHawkTemp},\cite{GibbHawkEntropy}. Notice too that the semiclassical regime yields as saddle point of the euclidean 
path integral of gravity the inverse value of $\Lambda$ \cite{HawkPhysLett}: 
\begin{equation} \label{saddlepoint}
3 m_P^2 /\Lambda:\qquad\; \mbox{saddle point of quantum gravity path integral}
\end{equation}
This expression is precisely our quantum dual cosmological constant $\Lambda_Q$: 
\begin{equation} \label{LambdaQ}
\Lambda_Q = \lambda_P/\Lambda = 3h_P^2/\Lambda
\end{equation}
The reason why the saddle point of the Euclidean path integral of gravity is the inverse 
of $\Lambda$ is simply because the cosmological constant acts in the gravitational action 
as a Lagrange multiplier as it is only coupled to the space-time volume of the Universe,  
implying the term $\Lambda L^4_\Lambda$. The quantum gravity context and the semiclassical regime 
in which Eq.(\ref{saddlepoint}) does appear 
show that the classical/semiclassical gravitational entropy $S_{\Lambda}$ and the classical and 
quantum temperatures $T_{\Lambda}$, $T_Q$ are completely in agreement with the physical 
context of the classical-quantum duality including gravity in which we describe it.

\begin{itemize}
\item{The Universe at its present age $H$, is in a {\it classical gravitational state or regime} of classical 
radius $L_H = c/H$ and classical cosmological constant $\Lambda = 3 H^2$. In Planck units, the 
gravitational entropy $S_{\Lambda}$ of the Universe today, and thus the  {\it classical gravitational entropy}, is {\it precisely} the inverse of the today cosmological constant value, ie 
\begin{equation} \label{SSLambdavalue}
S_{\Lambda} /s_P = (L_H/l_P)^2 = (h_P/H)^2 = (\lambda_P / \Lambda) = 10^{122}
\end{equation}  
This is precisely the inverse of the today classical $\Lambda$ density $\rho_{\Lambda}$ 
in Planck units $\rho_P$: $ \rho_P /\rho_{\Lambda} =  10^{122}$.}
\item{{\bf The $\Lambda$ density} $ \rho_{\Lambda} /\rho_P = 10^{-122} = \Lambda / \lambda_P$ observed today 
is {\bf precisely the quantum entropy} $S_Q / s_P =  \rho_{\Lambda} /\rho_P$, namely the area of the 
Universe of quantum radius $L_Q = l_p^2/ L_H$, ie the quantum dual radius of $L_H$, which is the 
Compton radius $L_Q = \hbar /(c M_H)$ of the Universe of mass  $M_H =  L_H c ^2 /G$. 
That is to say: 
\begin{equation} \label{SSQvalue}
S_Q / s_P = s_P/ S_H = (\Lambda / \lambda_P) = 10^{-122}.
\end{equation} 
The quantum gravitational entropy $S_Q$ is {\it precisely} the quantum dual of the classical gravitational entropy $S_\Lambda$ through its Planck scale value $s_P$:  
\begin{equation} \label{SrhoQLambda}
S_{\Lambda} = s_P \;\left(\frac{\rho_Q}{\rho_P}\right) = s_P \;\left(\frac{\lambda_P}{\Lambda}\right) = s_P \; 10^{+122}
\end{equation}

\begin{equation} \label{SQrhoLambda}
S_Q = s_P \; \left(\frac{\rho_\Lambda}{\rho_P}\right) = s_P \; \left(\frac{\Lambda}{\lambda_P}\right) = s_P \; 10^{-122}
\end{equation} 
The {\it total} $Q\Lambda$ gravitational entropy turns out the sum of the three components as it must be: 
classical (subscript $\Lambda$), quantum dual (subcript $Q$) and Planck value (subscript P) corresponding 
to the tree gravity regimes:
\begin{equation} \label{Stotalvalue}
S_{Q\Lambda} = 2 \;[s_P + \frac{1}{2}(S_{\Lambda} + S_Q)]
= 2 \;s_P \;[\; 1 + \frac{1}{2}(10^{+122} + 10^{-122})\;]
\end{equation}}
\item{The gravitational entropy $S_{\Lambda}$ of the present time large {\it classical Universe} 
is a very {\it huge number}, 
consistent with the fact that the Universe today contains a very huge amount of information. 
In order for $S_{\Lambda}$ to be associated with a {\it vacuum} energy density this must 
be a {\it very high density}: This is precisely the 
{\it quantum vacuum density $\rho_Q$ or quantum cosmological constant $\Lambda_Q$, 
 which are the quantum duals -quantum precursors- of the classical density $\rho_{\Lambda}$ and classical cosmological 
 constant $\Lambda$} respectively.}
\item{The value of $\Lambda$ today, that is the {\it classical} cosmological constant value, as a 
classical vacuum energy density $\rho_\Lambda$ is {\it naturally} a very small value because the 
accelerated Universe {\it today} is a {\it classical} and 
{\it dilute vacuum} Universe (in contrast to the quantum and highly dense super-Planckian very early 
state of the Universe). This is consistent with the well established set of observational results 
(refs \cite{VoidsHistory},\cite{Voids1} and refs therein) showing that the Universe today is an 
{\it empty} Universe dominated by {\it voids and supervoids}: Observations, numerical simulations and analytic 
results agree in the distribution of voids and supervoids which are the large scale vacuum sites of dominance of dark energy, 
(see for ex refs \cite{VoidsHistory},\cite{Voids1},\cite{VoidsPRL} and refs therein).} 
\item{On the contrary, the  quantum particle physics vacuum energy is the {\it quantum} dual density $
\rho_Q$ which is a huge value $10^{122}\;m_P$ deep inside in the quantum super-Planckian precursor era 
within a extremely small quantum radius $ L_Q$. The density $\rho_Q$ is the quantum 
dual of $\rho_\Lambda$ through its Planck scale value $\rho_P$:
\begin{equation} \label{rhoQ2}
\rho_Q = \frac{\rho^2_P}{\rho_\Lambda} = \rho_P \left(\frac{L_\Lambda}{l_P}\right)^2 = 
\rho_P \left(\frac{\lambda_P}{\Lambda}\right)
\end{equation}
The two densities, $\rho_{\Lambda}$, and $\rho_{Q}$, are {\it the same concept}: the vacuum energy density, {\it in two 
different states} (early superplanckian quantum phase and present time classical stage) of the Universe. 
The two densities are components of the {\it complete} $\rho_{Q \Lambda}$ density.}
\item{The {\it complete} $Q\Lambda$ density (classical plus quantum density) is :
\begin{equation} \label{rho3}
\rho_{Q \Lambda} = \rho_\Lambda + \rho_Q + \rho_P = \rho_P \left(\frac{\rho_\Lambda}{\rho_P} +  \frac{\rho_P}{\rho_
\Lambda} + 1 \right)  = \lambda_P \left(\frac{\Lambda}{\lambda_P} +  
 \frac{\lambda_P}{\Lambda} + 1 \right)
\end{equation}
In the case of the Universe till {\it today}, these values are:
\begin{equation} \label{rhototalvalue}
\rho_{Q \Lambda} = \rho_P\;(\;10^{-122}\; + \;10^{122}\; + 1 \;)
\end{equation}}
\end{itemize}
{\bf Summing up:}  The present Universe today of large classical horizon radius $L_\Lambda$ and very 
low density $\rho_\Lambda$
is a empty or dilute vacuum Universe (dominated by voids and supervoids) and {\it not} a dense quantum Universe. 
The very early Universe is a highly quantum dense Universe. The classical dilute Universe today and the highly dense very early quantum super-Planckian Universe are classical-quantum duals of each other in the precise meaning of the classical-quantum duality including gravity:

\medskip

The classical Universe today $U_{\Lambda}$ is clearly characterized by the set of physical 
magnitudes or observables (size/age,  mass, density, temperature, entropy):
$U_{\Lambda} \equiv (L_\Lambda, M_\Lambda, \rho_\Lambda, T_\Lambda, S_\Lambda)$. 
The highly dense very early quantum Universe $U_Q$ is characterized by the corresponding set of quantum dual physical magnitudes $U_Q \equiv (L_Q, M_Q, \rho_Q, T_Q, S_Q)$ in the precise meaning of the classical-quantum duality:
\begin{equation} \label{Udual}
U_Q =  \frac{u_P^2}{U_\Lambda}, \qquad u_P \equiv (l_P, m_P, \rho_P, t_P, s_P)
\end{equation} 
The {\it total} Universe is composed by their classical/semiclassical and quantum phases:
\begin{equation} \label{Utotal}
U_{Q\Lambda} = \left(\; U_Q  +  U_{\Lambda} + u_P \;\right)
\end{equation}
The Universe at its {\it present age} is a {\it classical} Universe of huge classical radius $L_\Lambda$ 
and thus a huge {\it classical horizon area} $ A_\Lambda$
and so a huge value for the classical/semiclassical gravitational entropy $S_\Lambda = 10^{+122}\; \pi \kappa_B$.  The entropy $S_\Lambda$ is related 
to the {\it classical} area $ A_\Lambda \approx L_{\Lambda}^2 \approx 1/\Lambda$ and thus to the inverse of the classical $\Lambda$. 
This explains {\it why} the cosmological constant has such a small value and $S_\Lambda$ a so high one.
$S_\Lambda$ today is {\it not} proportional to $\rho_\Lambda$ which is a extremely small value, but to the {\it quantum dual} of $\rho_\Lambda$, ie  the quantum density $\rho_Q$ which is its precursor: a extremely high (superplanckian) value in the extreme past. This is clearly seen 
from Eqs.(\ref{SSLambdavalue}),(\ref{SrhoQLambda}),(\ref{SQrhoLambda})   simply summarized as:
\begin{equation} \label{S3}
\frac{S_\Lambda}{s_P} = \left(\frac{L_\Lambda}{l_P}\right)^2 =  \frac{\rho_Q}{\rho_P} = 
\frac{\rho_P}{\rho_\Lambda} = \frac{\lambda_P}{\Lambda} = \frac{s_P}{S_Q} = 10^{122}
\end{equation} 

\medskip

By going back in time along the Universe evolution, from the present era today to the early Universe stages, the gravitational entropy of the Universe is decreasing from its present huge value $S_\Lambda = 10^{122}\pi\kappa_B$ today at the age 
$10^{61} t_P$ to its inflationary value $S_\Lambda = 10^{12}\pi\kappa_B$ during inflation at the time $10^{-6} t_P$, then falling to its Planck value $s_P = \pi\kappa_B$ at the Planck time $t_P$ and then following decreasing till reaching its extreme lowest known value $ S_Q = 10^{-122}\pi\kappa_B$  in the earliest quantum era at $10^{-61} t_P$. 

\medskip

The extreme smallest value of the entropy is the quantum dual of the largest known classical entropy at the horizon today:
\begin{equation} \label{Svalue}
\frac{S_\Lambda}{s_P} = \frac{s_P}{S_Q} =   10^{+ 122}.                      
\end{equation} 

\medskip

The largest time and length in the Universe are its present age and horizon size:  
$10^{+61}\; t_P$ and $L_\Lambda = 10^{+61}\; l_P$ . The smallest time and length in the Universe 
are the quantum duals of them: $10^{-61}\; t_P$ and $L_Q = 10^{-61}\; l_P$  

\medskip

{\bf The classical and quantum $\Lambda$ Temperatures.} The above results can be also seen in terms of the classical and quantum temperatures of the Universe 
$T_\Lambda$ and $T_Q$. Eqs (\ref{TH2}),(\ref{TQ2}) and Eq.(\ref{Udual}) for the classical today Universe 
$U_{\Lambda}$ and its quantum earlier dual $U_{Q}$ yield 
the following enlighting summary :
\begin{equation}
\frac{T_\Lambda}{t_P} = \left(\frac{L_\Lambda}{l_P}\right) = 
\frac{h_P}{H} = \sqrt{\frac{\lambda_P}{\Lambda}} = \sqrt{\frac{S_\Lambda}{s_P}}
= \sqrt{\frac{s_P}{S_Q}} = 10^{61}
\end{equation}
\begin{equation}
\frac{T_Q}{t_P} = \left(\frac{l_P}{L_\Lambda}\right) = 
\frac{H}{h_P} = \sqrt{\frac{\Lambda}{\lambda_P}} = \sqrt{\frac{S_Q}{s_P}} = 
\sqrt{\frac{s_P}{S_\Lambda}} = 10^{-61}
\end{equation}

\medskip

From the above results and the observed value of $\Lambda$ today Eq.(\ref{Lambdavalue}), the values of the classical 
temperature $T_{\Lambda}$ of the Universe {\it 
today}, and the temperature $T_Q$ of its quantum precursor are: 
\begin{equation}
T_{\Lambda \; today} = t_P \;\sqrt{\frac{\lambda_P}{\Lambda}} =  (0.5875 \pm 0.0800) \; 10^
{61} \; t_P
\end{equation}
\begin{equation}
T_{Q \; today} = t_P \;\sqrt{\frac{\Lambda}{\lambda_P}} = (1.6865 \pm 0.0229)\; 10^{-61} \; t_P        
\end{equation}
That is:
\begin{equation}
 T_{\Lambda \; today} = (0.5875 \pm 0.0800) \; 10^{93} K
\end{equation}
\begin{equation}
T_{Q \; today}= (1.6865 \pm 0.0229)\; 10^{-29} K 
\end
{equation} 

The {\it total} or {\it complete} $Q\Lambda$ Temperature $T_{Q\Lambda}$ 
Eq.(\ref{TQH2}) is precisely {\it the sum} of the different components (classical plus quantum):
\begin{equation} \label{Ttotal}
T_{Q\Lambda \; today} =  [\; T_\Lambda + T_Q + t_P \;]_{today}  = (10^{61} + 10^{-61} + 1)\; t_P
\end{equation}

In the classical large Universe today, the classical {\it today} component $T_\Lambda$ dominates, 
as it must be. In its quantum precursor, the quantum Planck component $t_P$ dominates, as it must be.

\bigskip

{\bf Comparison to the Inflation $\Lambda$ Temperatures and Entropies}: 
For comparison, the temperatures $T_\Lambda$ and $T_Q$ for the Inflation era are:
\begin{equation}
T_{\Lambda \;inflation} = t_P \;\sqrt{\frac{\lambda_P}{\Lambda}}_{inflation} = 10^{38} K 
=  10^{6} \; t_P 
\end{equation}
\begin{equation}
T_{Q \;inflation} = t_P \;\sqrt{\frac{\Lambda}{\lambda_P}}_{inflation} = 10^{26} K 
= 10^{-6}\; t_P
\end{equation}
In the classical inflation era, the classical $T_\Lambda$ component dominates as it must be, while in its quantum precursor era, $t_P $ dominates as it must be.  The complete $T_{Q \Lambda \;inflation}$ is the sum of its components, 
\begin{equation} \label{Tinftotal}
T_{Q\Lambda \;inflation} =  [ \; T_\Lambda + T_Q + t_P \;]_{inflation} = (10^{6} + 10^{-6} + 1)\; t_P
\end{equation}
Eqs.(\ref{Ttotal}) and (\ref{Tinftotal}) show consistently that the difference  between 
the classical and quantum temperatures
(which is huge in the today classical Universe highly dominated by the classical $T_
\Lambda$) diminishes 
in the early and more quantum stages as in Inflation where the difference 
between the two values $T_\Lambda = 10^{6}\; t_P$ and $T_Q = 10^{-6}\; t_P$ is considerably smaller than in the present time. 

\medskip

In addition, Eq (\ref{Tinftotal})  {\it consistently} reflects the {\it semi-classical or  semi-quantum gravity} character of Inflation. 
In other words, as well as the Planck scale $ m_P$ is from the classical side the crossing to the quantum 
gravity regime, the {\it Inflation scale $10^{-6} m_P$ in the classical phase is the typical scale for 
the semi-classical gravity} regime. And the quantum dual Inflation scale in the quantum precursor phase is 
consistently $10^{6} m_P$. (This last could be viewed as a "semi-quantum gravity" scale, "low" with respect 
to the higher superplanckian scales of the earlier quantum stages, the highest $H = 10^{61} hp$ being at the extreme quantum past. Whatever be, classical or quantum, Inflation is at $10^{\pm 6}$ from the Planck scale). Consistently, this can be also seen in terms of the classical and quantum entropies $S_{\Lambda}$ and 
$S_Q$ of Inflation: 
\begin{equation}
S_{\Lambda \;Inflation} = s_P \;\left(\frac{\lambda_P}{\Lambda}\right)  
= 10^{+12} \;s_P  = 10^{+12} \; \pi \; \kappa_B  
\end{equation}
\begin{equation}
S_{Q \;Inflation} = s_P \; \left(\frac{\Lambda}{\lambda_P}\right) = 10^{-12} \;s_P  
=  10^{-12} \; \pi \; \kappa_B
\end{equation}

$S_{\Lambda \;Inflation}$ in the classical Inflation stage at $10^{6}t_P$ is larger than its precursor value 
$S_{Q \;Inflation}$ in the quantum Inflation precursor stage, (arrow of time), as it must be. 

\bigskip

{\bf Cosmological Constant Summary:}
Summing up, the solution to the cosmological constant can be explicitely summarized in the following simple equations:
\begin{equation} \label{Lambda1}
 \frac{\lambda_P}{\Lambda_Q} = \frac{\Lambda}{\lambda_P} =
\frac{\rho_\Lambda}{\rho_P} = \frac{S_Q}{s_P} = \left(\frac{T_Q}{t_P}\right)^2 = 10^{-122}
\end{equation}
and
\begin{equation} \label{LambdaQ2}
\frac{\Lambda_Q}{\lambda_P}  = \frac{\lambda_P}{\Lambda} =
\frac{\rho_Q}{\rho_P} = \frac{S_\Lambda}{s_P} = \left(\frac{T_{\Lambda}}{t_P}\right)^2 = 
10^{+122} 
\end{equation}

The  {\it complete} $Q\Lambda$ cosmological constant $\Lambda_{Q\Lambda}$ or complete vacuum energy density 
$\rho_{ Q\Lambda}$ is given by:
\begin{equation}  \label{LambdaQLambdavalue}
\Lambda_{Q \Lambda} = \Lambda + \Lambda_Q + \lambda_P = \lambda_P \left (\; 
\frac{\Lambda}{\lambda_P} + \frac{\lambda_P}{\Lambda} + 1 \;\right)
 = \lambda_P \;(\; 10^{-122} + 10^{+122} + 1 \;)
\end{equation}
which is the sum of its classical and quantum {\it components}.

\medskip

The observed value today is the classical $\Lambda $ vacuum value $10^{-122}$ corresponding to the classical 
Universe today which is a large, classical and empty or vacuum  dilute Universe. This is the 
main physical reason for such low value. The computed particle physics quantum $\Lambda_Q$ value  
$10^{+122}$ is the vacuum value corresponding to the very early 
Universe  which is a  extremely small, quantum and high density (superplanckian) vacuum. This is the 
main physical reason for such high value.

\medskip

All physical magnitudes: the vacuum energy density, the cosmological constant,  
the gravitational entropy and gravitational temperature, both classical and quantum are linked by the 
classical-quantum (or wave-particle) duality through the Planck scale.

\medskip

Eqs. (\ref{Lambda1}),(\ref{LambdaQ2}),(\ref{LambdaQLambdavalue}) 
concisely and synthetically express such classical-quantum dual relations and {\it explain  why} 
the classical vacuum cosmological constant $\Lambda$ or classical density $\rho_\Lambda$ {\it coincides} 
with such observed {\it low value} $10^{-122}$ in Planck units. 
The vacuum computed density from particle physics  $10^{+122}$ is a quantum extreme vacuum value, 
it is precisely the {\it quantum dual} density $\rho_Q$ to the classical density $\rho_ \Lambda$ today. 
 
\medskip

{\bf Cosmological Constant Conclusion:} The quantum vacuum density or quantum $\Lambda_Q =\rho_ Q = 10^{+122}$ (in Planck units) {\it is  not} what is 
observed today and must be consistently such way because the Universe today {\it is  not} in a quantum 
super-planckian gravitational state. The Universe today is in a classical gravitational regime and dilute classical state. And what is observed today is consistently and correctly the 
classical low dilute value $\rho_ \Lambda = 10^{-122}$ or classical vacuum $\Lambda$ corresponding to the 
classical Universe today.  The quantum vacuum density $\rho_ Q = 10^{+122}$ {\it is } a quantum 
super-planckian value and must be consistently such way because it is the quantum precursor in a quantum gravitational  super-planckian very early past state. The past Universe before the Planck time is in a quantum gravitational super-planckian regime and highly quantum superplanckian state, precursor of the  observed  era today of the Universe. 

\bigskip

{\bf The Whole History. An Unifying Picture:}
We see that going back in time along the Universe evolution from the present era to the early stages where the Universe 
becames more and more quantum, the classical temperature $T_{\Lambda}$ decreases, as it 
must be, the quantum temperature $T_Q$ becomes higher and the values of the Classical and 
Quantum temperatures $T_{\Lambda}$ and $T_Q $ Eqs.(\ref{TH2}),(\ref{TQ2}) become closer of each other, the 
difference dissapearing at the {\it Planck scale}:
$ T_{\Lambda} = T_Q = t_P$, which is the {\it crossing scale} between the classical/semiclassical 
and quantum gravity regimes or eras. 

\medskip

Similarly, going back in time, from the present era to the early quantum eras of the Universe, the classical gravitational entropy $S_{\Lambda}$ decreases from its huge 
value today $S_{\Lambda \;today} = 10^{122}\;s_P$ at $10^{61} t_P$ to the inflationary value $S_{\Lambda 
\;inflation} = 10^{12}\;s_P$ in the Inflation era (semiclassical gravity era) at $10^{6} t_P$, then after 
descending to its small Planckian value $s_P = \pi\kappa_B$ at the Planck time $t_P$, in which 
it enters the quantum and super-Planckian regime, decreasing for instance to 
$S_{Q\; inflation} = 10^{-12}\;s_P$ (the quantum dual phase of Inflation) at the time $ 10^{-6} t_P$, untill 
 reaching its smallest extreme value $s_P \; 10^{-122}$ at $ 10^{-61}\; t_P$. 
\begin{figure}\centerline{\includegraphics[height=20 cm,width=26 cm]{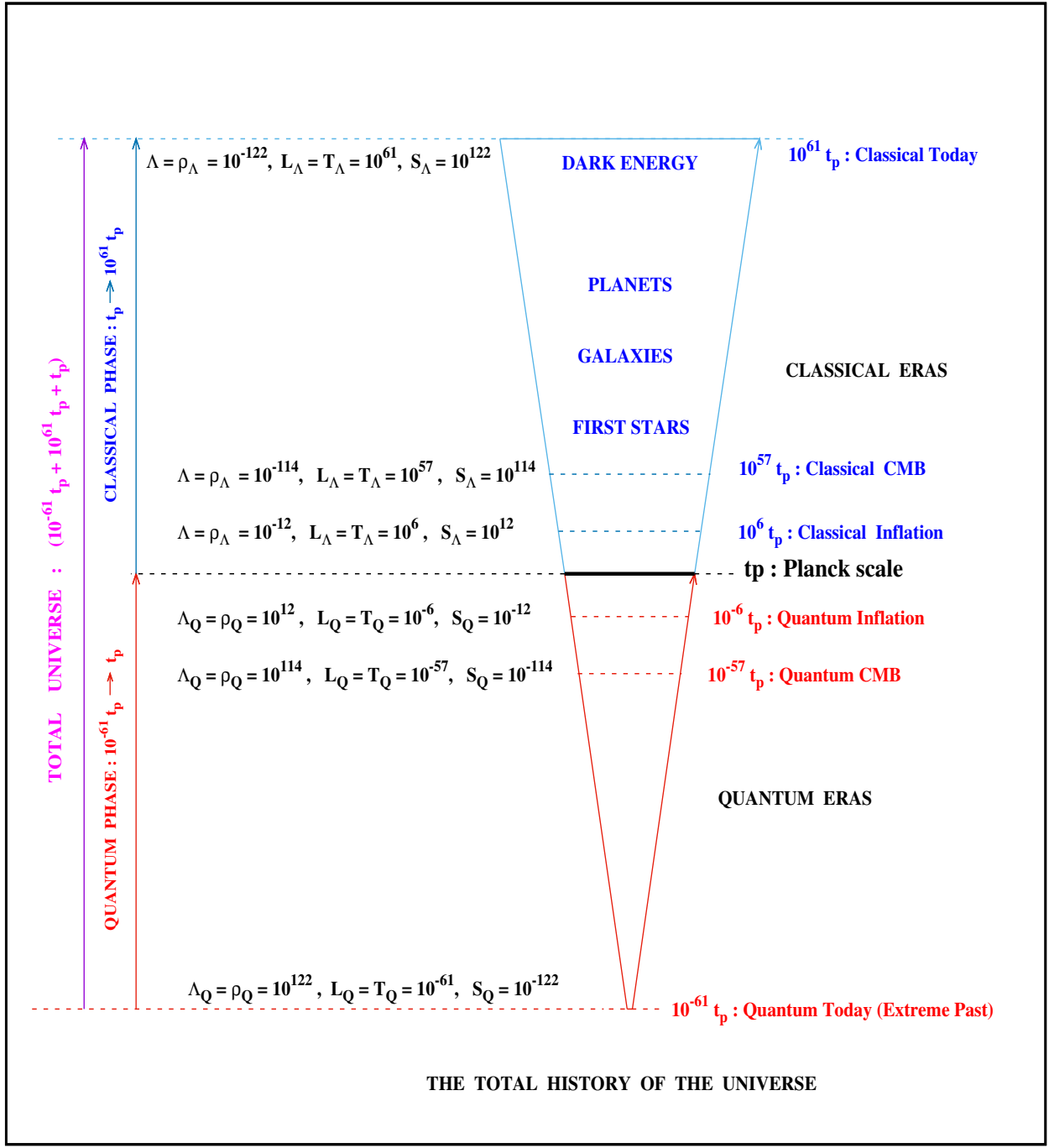}}
\setlength\abovecaptionskip{0 cm}\caption{\setlength{\baselineskip}{0.7\baselineskip}
{\bf The Standard Model of the Universe 
completed by Quantum physics in terms of its Gravity 
history.} The Universe is composed of two 
big phases after and before the Planck scale $t_P$ (the {\it crossing scale}). The post-planckian classical gravitation phase is the Universe from $ t_P $ to the 
present age $10^{61} t_P$. The quantum (planckian and super-planckian) phase
from the extreme past $10^{-61} t_P$ to $t_P$ is its precursor. 
The complete history goes from $10^{-61} t_P$ to $10^{61} t_P$. Quantum physics, Planck scale, natural to the system, and gravitation unify and clarify the whole history.  See the text at the end of Section XI and the complete figure caption there.}
\end{figure}

\medskip

Conversely, starting from the earliest past quantum era from $10^{-61}\; t_P$ to $t_P$, the quantum 
entropy $S_Q$ increases from its 
extreme small value $S_{Q} = 10^{-122} \;s_P$ at the earliest time $ 10^{-61}\; t_P$ 
till for instance its quantum inflation value $10^{-12} \;s_P$ at the time $10^{-6}\; t_P$, to its Planck 
value $S_Q = s_P = \pi \kappa_B$ at the Planck time $t_P$, the {\it crossing scale}, after which it 
goes to its semi-classical inflationary  value $S_{\Lambda \;inflation} = 10^{12} \;s_P$ at 
the classical inflationary stage at $10^{6} \; t_P$ 
and it follows {\it increasing and classicalizes} till the extreme maximal 
classical value today $S_{\Lambda} = 10^{122} \;s_P$ at the present Universe time $10^{61}\; t_P$.  
And $S_{\Lambda}$ will continue increasing to higher values in the future as far as the 
Universe will continue expanding its horizon.

The {\it total} $Q\Lambda$ gravitational entropy (for the whole history) is the sum of the three values 
above discussed 
corresponding to the three regimes: classical $\Lambda$, quantum dual $Q$ and Planck 
values (subcript $P$), Eq.(\ref{LambdaQLambdavalue}).
In the past remote and more quantum (Q) eras: $10^{-61} \;t_P \leq  t \leq t_P$, 
the Planck entropy value $s_P = \pi \kappa_B$ dominates $S_Q$. In the classical eras:
$t_P \leq  t \leq 10^{61} t_P$, the today entropy value $S_\Lambda = 10^{+122} s_P$  
dominates.

{\bf The whole picture is depicted in Figure (1)}, where: $\Lambda$ refers to the cosmological constant (or associated Hubble-Lemaitre constant H) in the 
Classical gravity phase. Q means quantum. P means Planck scale. Planck's units, natural to the 
system, greatly simplify the history. (The complete history is a theory of pure numbers). Each stage is characterized by the set of main physical gravitational  quantities: ($\Lambda$, density $\rho_\Lambda$, 
size $L_\Lambda$, gravitational temperature $T_\Lambda$ and entropy $S_\Lambda$). In the Quantum phase, 
their corresponding  quantum precursors are labeled with the subscript Q. Classical and 
quantum precursor stages and their associated physical quantities are classical-quantum duals of each other in 
the precise meaning of the classical-quantum or wave-particle duality including gravity Eq.(\ref{Udual}). Total means the whole history including the two phases or regimes. The present age of the Universe 
$10^{61}$, (with $\Lambda = \rho_\Lambda = 10^{-122} = 1/S_\Lambda$) is a {\it lower bound} to the 
future Universe age and similarly for the present entropy value $S_\Lambda$. While 
$10^{-61}$, (with $\Lambda_Q = 10^{122} = \rho_Q = 1/S_Q$ 
is an {\it upper bound} to the extreme past (origin) of the Universe and quantum initial entropy, (arrow of time).  
[Similarly, the values given in Fig.1 (in Planck units) for the CMB are the classical CMB age ($3.8 \; 10^5 yr = 10^{57} t_P$) and the set of caracteristic gravitational properties of the Universe at this age, and their corresponding quantum precursors in the quantum  preceding era at $10^{-57} t_P$. $T_\Lambda$ and $S_\Lambda$ are also un upper bound to the temperature and entropy of the CMB photon radiation.]

\section{Conclusions}

We have accounted in the Introduction and along the paper 
the main new features of the paper and will not include all of them here.
We refer to Section I for a summary of the main results and the end of 
previous Section XI for the whole picture.
\begin{itemize}
\item{
We described classical, semiclassical and quantum de Sitter regimes. 
A clear picture for the de Sitter background and the whole Universe epochs emerges, going
beyond the current picture, both for its classical and quantum regimes, depicted in Fig (1).

This is achieved by recognizing the relevant scales of the classical and quantum regimes of gravity.
They turn out to be the classical-quantum duals of each other, in
the precise sense of the wave-particle (de Broglie, Compton) duality extended to the quantum gravity (Planck and super-Planck) domain: wave-particle-gravity duality.}
\item{
Concepts as the Hawking temperature and the usual (mass) temperature 
are shown to be precisely the same concept
in the different classical and quantum gravity regimes respectively.  
Similarly, it holds for the Bekenstein-Gibbons and Hawking entropy.}
\item{
An unifying clarifying picture has been provided 
including the main physical gravitational intrinsic magnitudes of the Universe: 
age, size, mass, vacuum density, temperature, entropy, 
in terms of the cosmological constant covering the relevant gravity regimes or 
cosmological stages: classical, semiclassical and 
quantum -planckian and superplanckian- eras.} 
\item{
Cosmological evolution goes from a super-planckian and planckian quantum phase to a semiclassical accelerated de Sitter era (field theory inflation), then to the classical phase untill the present de Sitter era. The wave-particle-gravity duality precisely manifests
in this evolution, between the different gravity regimes, and could be view 
as a mapping between asymptotic (in and out) states characterized by the sets
$U_\Lambda$ (or $U_H$) and $U_Q$, and thus as a Scattering-matrix description: The most early 
quantum super-Planckian state 
in the remote past being the in-state, and the very late classical dilute state
being the far future or today out-state.} 
\item{
Along its physical history, from the very early stages to the present time, 
the Universe evolved from quantum stages to classical physics stages: that is to say, the Universe {\it 
classicalized}.  And conversely, from the present time to the earlier stages, the 
Universe becomes {\it quantized}.
Inflation is part of the standard cosmological model and is supported by the CMB data 
of temperature and temperature-E polarisation anisotropies. This points to $10^{-6} m_P
$, (or $10^{-5} M_P$ for the reduced mass $M_P = m_P/\sqrt{8 \pi}$) as the energy scale of Inflation \cite{CiridVS},\cite{BDdVS}, safely below the Planck energy scale $m_P$ of the onset of quantum gravity. This implies that Inflation is consistently in the {\it semiclassical gravity regime}. This 
in turn implies that the preceding phase of Inflation corresponds to a quantum gravity 
phase in the Planckian and super-Plankian quantum gravity domaine. Inflation being a de Sitter, (or quasi de Sitter) stage, it has a smooth space-time curvature {\it without any physical space-time singularity}.}
\item{
Integrating the above different pieces of knowlodge, and because the more earlier known stages of the Universe are de Sitter (or quasi de Sitter) eras, it appears as a consequence of our results that there is {\bf no singularity} at the Universe's origin. First: the so called $t = 0$ Friedman-Robertson Walker  mathematical singularity is {\bf not} physical: it is the result of the extrapolation without any quantum physics of the purely
classical (non quantum) General Relativity theory, out of its domain of physical validity. The Planck scale is not merely a useful system of units but a physically meaningful scale: quantum gravity. The Planck scale precludes the extrapolation to zero time or length. This is precisely what is expected from quantum physics in gravity: the smoothness of the classical gravitational 
singularities. Second: Inflation (classical or quantum) in the very past 
($10^{6} t_P$ or $10^{-6} t_P$ is mainly a de Sitter (or quasi de Sitter) smooth constant 
curvature era {\it without any curvature singularity}. Third: the extreme past 
 (at $10^{-61} t_P$) is a superplanckian de Sitter state of high {\it bounded} superplanckian constant curvature and therefore {\it without singularity}. Of course, this paper is not devoted to the singularity issue  but this argument and the whole picture emerging from this paper indicate the trend and insight into the problem.}
\item{
The main property used here is the classical-quantum duality, which is a 
universal foundational milestone of quantum theory.
Further couplings, interactions and background fields can be added. The conceptual
results here will not change by adding further couplings or interactions,
or further background fields to the background here. Of course, this is just a first input in the 
construction of a complete physical theory and understanding {\it in agreement with observations}.}
\item{
The existence and present state of the Universe is physically 
explained because of the classical-quantum duality as a basic and universal
property of Nature: Our known classical Universe does exist 
{\it precisely} because there existed a preceding phase or precursor:  Such preceding phase is 
{\it exactly} the quantum dual phase of the existing known classical phase, the Planck time being 
precisely {\it the crossing} time between the two phases, as given by Eq.(\ref{Udual}). 
The Planck time is the transition to the Classical/semiclassical gravity Universe from the "end" ("late" time or entropy) of the Quantum dual preceding phase.}
\item{
Besides its conceptual and fundamental physics interest,
this framework revealed of deep and useful clarification for relevant cosmological eras and its 
quantum precusors and for problems as the cosmological constant. This could provide realist insights 
and science directions where to place the theoretical effort for cosmological missions 
and future surveys such as Euclid, DESI and WFIRST for instance,
\cite{Euclid}, \cite{DESI}, \cite{WFIRST},
 and for the searching of cosmological quantum gravitational signals.}
\item{
The exhibit of $(c, G, h)$ helps in recognizing the different relevant scales and physical regimes.  
Even if a hypothetical underlying "theory of everything" could only require
pure numbers (option three in \cite{Duff}), physical touch
at some level asks for the use of fundamental constants 
\cite{Okun},\cite{Sanchez2003}. Here we used
three fundamental constants, (tension being $c^2/G$). It appears from our study   
here and in ref \cite{Sanchez2019}, that a complete quantum theory 
of gravity would be a theory of pure numbers.}
\end{itemize}

\bigskip

{\bf ACKNOWLEDGEMENTS}

\bigskip

The author acknowledges the French National Center of Scientific Research (CNRS)
for Emeritus Director of Research contract and F. Sevre for help with the figure. 
This work was performed in LERMA-CNRS-Observatoire de Paris-
PSL University-Sorbonne University.

\end{document}